\documentclass[prstab,12pt,superscriptaddress,nofootinbib,showkeys,showpacs]{revtex4}
\usepackage{times,latexsym,epsfig,epsf,graphicx,amsmath}

\usepackage{amssymb}
\usepackage{hyperref} 
\usepackage{dcolumn}%

\begin{document}

\newcommand{\cracow1}{\affiliation{AGH University of Science and Technology, Krakow, Poland}}
\newcommand{\cracow}{\affiliation{Cracow University of Technology, Warszawska 24 St., 31-155 Krakow, Poland}}
\newcommand{\ral}{\affiliation{STFC Rutherford Appleton Laboratory, OX11 0QX Didcot, UK}}
\newcommand{\saclay}{\affiliation{Irfu, CEA-Saclay, 91191 Gif-sur-Yvette, France}}
\newcommand{\iphc}{\affiliation{IPHC, Universit\'e de Strasbourg, CNRS/IN2P3, F-67037 Strasbourg, France}}





\title{ The SPL-based Neutrino Super Beam
}


\author{E.~Baussan}\iphc
\author{J.~Bielski}\cracow
\author{C.~Bobeth}\iphc
\author{ E.~Bouquerel}\iphc
\author{O.~Caretta}\ral
\author{P.~Cupial}\cracow1
\author{T.~Davenne}\ral
\author{C.~Densham}\ral
\author{M.~Dracos}\iphc
\author{M.~Fitton}\ral
\author{G.~Gaudiot}\iphc
\author{M.~Kozien}\cracow
\author{L.~Lacny}\cracow
\author{B.~Lepers}\iphc
\author{A.~Longhin}\saclay  
\author{P.~Loveridge}\ral
\author{F.~Osswald}\iphc
\author{P.~Poussot}\iphc
\author{M.~Rooney}\ral
\author{B.~Skoczen}\cracow
\author{B.~Szybinski}\cracow
\author{A.~Ustrzycka}\cracow
\author{N.~Vassilopoulos}\iphc
\author{D.~Wilcox}\ral
\author{A.~Wroblewski}\cracow
\author{J.~Wurtz}\iphc
\author{ V.~Zeter}\iphc
\author{M.~Zito}\saclay





\begin{abstract}
The EUROnu Super Beam work package has studied a neutrino beam based on SPL at CERN and aimed at MEMPHYS, a large water Cherenkov detector, proposed for the Laboratoire Souterrain de Modane (Fr\'ejus tunnel, France), with a baseline of 130 km. The aim of this proposed experiment is to study the CP violation in the neutrino sector. 

In the study reported here, we have developed the conceptual design of the neutrino beam, especially the target and the magnetic focusing device. Indeed, this beam present several unprecedented challenges, like the high primary proton beam power (4 MW), the high repetition rate (50 Hz) and the low energy of the protons (4.5 GeV). The design is completed by a study of all the main component of the system, starting from the transport system to guide the beam to the target up to the beam dump.

\end{abstract}



\maketitle


\newpage

\tableofcontents

\newpage 


\section{Introduction}
\label{sec:splsb}

The EUROnu Super Beam work package has studied the project of a neutrino beam based on the SPL at CERN and aimed at MEMPHYS, a large water Cherenkov detector, proposed for the Laboratoire Souterrain de Modane (Frejus, France), with a baseline of 130 km. The aim of this proposed experiment is to study the CP violation in the neutrino sector. 

In the study reported here, we have developed the conceptual design of the neutrino beam, especially the target and the magnetic focusing device. Indeed, this beam present several unprecedented challenges, like the high primary proton beam power (4 MW), the high repetition rate  (50 Hz) and the relatively low energy of the protons  (4.5 GeV). The design is completed by a study of all the main component of the system, starting from the transport system to guide the beam to the target up to the beam dump. 

The report is organized in the following way. In this section, we briefly present the overall system, with references to the previous studies and a summary of the main  parameters and dimensions. We then present the various components, the beam transport and distribution system (section \ref{sec:beamsw}), the target station (section \ref{sec:ts}), the target (section \ref{sec:target}), the horn (section \ref{sec:horn}).  Finally, the study of the activation and shielding of the system is presented in section \ref{sec:activation} and the optimization tool, neutrino fluxes and physics performances in section \ref{sec:optimization}.
This report presents only a summary of the main results obtained in the course of this study. A more complete description can be found in 
\cite{ref:wp2final} as well as in the various EUROnu  reports \cite{ref:wp2reports}.

First studies of this facility \cite{Ball:2000pj,Gilardoni:2004kr,Mezz03}
were performed assuming a 2.2~GeV proton beam and a
liquid mercury jet target associated with a 
single conic horn with a pulsed current of 300~kA. 
Later it was proposed \cite{Gilardoni:2003us} to supplement the system with an auxiliary horn 
(called reflector) enclosing concentrically the first horn and operated at 600~kA
 in order to focus also pions produced at larger angles. 
This scheme was adopted in \cite{Campagne:2004wt} and the horn shape re-optimized 
using the method described in \cite{Campagne:2004cd}. Further, the decay tunnel was re-optimized using different primary beam energies from 2.2 up to 8~GeV. Based on the neutrino fluxes of \cite{Campagne:2004wt} 
and an improved parametrization of the far detector, the physics performances of the project were presented 
in \cite{Campagne:2006yx} assuming a 3.5~GeV proton kinetic energy.

With respect to previous studies on this subject we propose a
new design based on the use of a solid target and a single magnetic horn 
operated with a lower value of the pulsed current (300-350~kA).
Such a setup simplifies the engineering complexity of the system 
avoiding difficult issues such as the containment of the mercury jet in a magnetic field free region,
the challenge of a power supply operating at 600~kA and 
the constraints related to mechanical stresses on the horn-reflector system induced by the high frequency current pulsing.

The proton beam for this facility will be provided by the high power SPL, followed by an accumulator ring. To reduce the challenge on the target and the horn system, in particular the heat to be removed, the stresses and the radiation damage, we have forseen a set of four identical target and horn units. Each target will then receive a full beam spill every 80 ms for a total power of 1 MW. 

\begin{figure}[htbp]
\begin{center}
\includegraphics*[width=7cm]{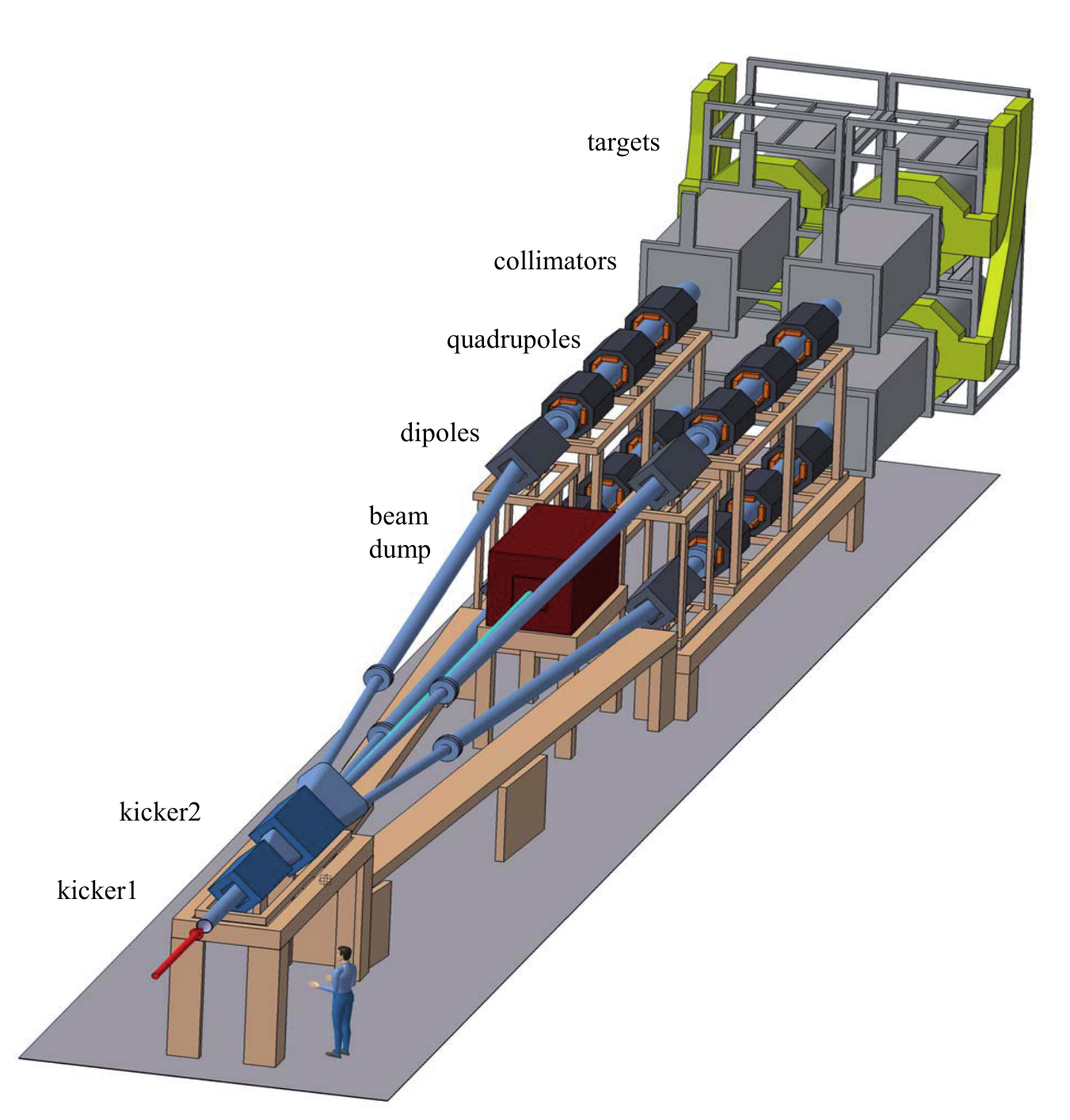}
\caption{The beam transport and distribution system.  }\label{fig:switch-overall}
\end{center}
\end{figure}

We present a view of the beam transport and distribution system in Fig.~\ref{fig:switch-overall}. The beam line, with a total length 30 m, is composed of two kickers, and then one dipole and three quadrupoles on each of the four separate transport lines.

\begin{figure}[htbp]
\begin{center}
\includegraphics[width=0.85\columnwidth]{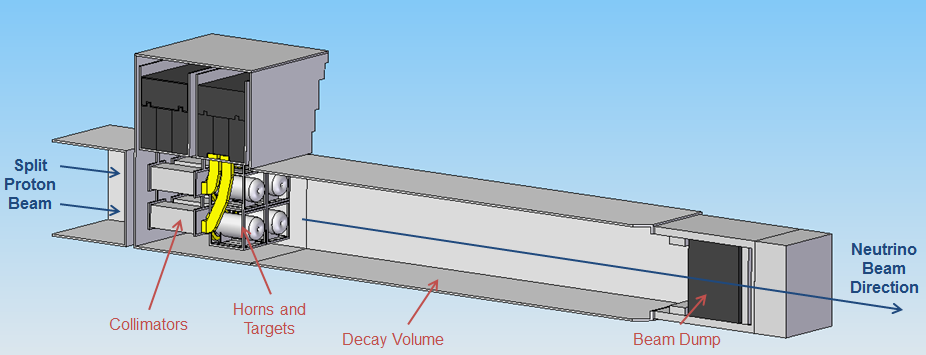}
\caption{Components of Target Station Beamline}
\label{FigTS:fig00}
\end{center}
\end{figure}

The target station is shown in Fig.~\ref{FigTS:fig00} and consists of the four targets and horns within a single large helium vessel. It is followed by the decay volume with a length of 25 m and by the beam dump. The thickness of the shielding around the decay volume is 2.2 m iron and 3.7 m concrete.  

The target (78 cm long and 2.4 cm diameter) is a made out of a titanium can filled with 3 mm diameter titanium spheres. It is cooled by a transversal helium flow. Each target is inserted inside a 2.5 m long magnetic horn, pulsed with a current of 300 kA. 

\section{The beam transport and distribution	}
\label{sec:beamsw}

\subsection{The superconducting proton linac (SPL) }

 The proton driver foreseen for the SPL-Super Beam is the High Power Super Conducting Proton Linac (HP-SPL) under study at CERN. The current design studies \cite{Bay-06,Bru-08} consider a beam power of 4 MW at 50 Hz repetition frequency with protons of about 4.5 GeV kinetic energy and a pulse duration of about 600~$\mu $s for neutrino physics applications. The parameters considered for the SPL in the latest study are reported in Table \ref{tab:HP}.

\begin{table}[h]
 \centering
\begin{tabular}{|p{2.3in}|p{0.9in}|} \hline
\textbf{Parameters} & \textbf{Value} \\ \hline
Energy & 4.5 GeV \\ \hline
Beam power & 4.0 MW \\ \hline
Rep. rate & 50 Hz \\ \hline
Average pulse current & 40 mA \\ \hline
Transverse emittances & 3$\pi $.mm.mrad \\ \hline
Beam width & 2 mm \\ \hline
Peak pulse current & 64 mA \\ \hline
Chopping ratio & 62 \% \\ \hline
Beam pulse length & 0.6 ms \\ \hline
Protons per pulse for PS2 & 1.5 x 10${}^{14}$ \\ \hline
Beam duty cycle & 2.0 \% \\ \hline
Number of klystrons (LEP) & 14 \\ \hline
Number of klystrons (704 MHz) & 57 \\ \hline
Peak RF power & 219 MW \\ \hline
Average power consumption & 38.5 MW \\ \hline
Cryogenics av. Power consumption & 4.5 MW \\ \hline
Cryogenic temperature & 2.0 K \\ \hline
Length & 534 m \\ \hline
\end{tabular}
 \caption{Parameters of the HP-SPL  \cite{Bay-06,Bru-08}.}\label{tab:HP}
\end{table}

\subsection{ The accumulator ring}

  The pulse duration of the proton beam delivered on the SPL-Super Beam target-horn station should be less than $5$ $\mu $s \cite{Bob-11,Mau-01} . 
 For this reason an additional accumulator ring is required interfacing the SPL and the target-horn station. 
 
 Dedicated design studies have been performed for the Neutrino Factory \cite{Aib-08,Ben-09} which requires a combination of accumulator and compressor rings in order to achieve a bunch length of 2 ns rms after compression. For the Super Beam the accumulator ring is sufficient and among the considered scenarios  the 6-bunch per pulse option is most suited allowing for the lowest values of the local power distribution inside the target. This scenario foresees 6 bunches per pulse with bunch length 120 ns and gaps of 60 ns. The longitudinal bunch profile has a trapezoidal shape with linear density rising and falling at the beginning and end within 10 ns in order to avoid longitudinal instabilities.


\begin{table}[h]
 \centering
\begin{tabular}{|p{2.5in}|p{0.9in}|} \hline
\textbf{Parameters} & \textbf{Value} \\ \hline
Energy & 4.5 GeV \\ \hline
Relativistic $\gamma $ & 6.32907 \\ \hline
Number of bunches & 6 \\ \hline
Beam size, $\sigma $ & 2 mm \\ \hline
Transverse emittances (rms) & 3$\pi $.mm.mrad \\ \hline
Total bunch length & 120 ns \\ \hline
RMS momentum spread (dp/p) & 0.863 x 10${}^{-3}$ \\ \hline
Circumference & 318.448 m \\ \hline
Average $\beta $ function ($\beta $${}_{x}$, $\beta $${}_{y}$) & 20,20 m \\ \hline
Momentum compaction, $\alpha $${}_{0}$ & 0.0249643 \\ \hline
Nominal tune, Q${}_{x}$/Q${}_{y}$ & 7.77,  7.77 \\ \hline
Natural chromaticity, Q'${}_{nat}$ & -8.4, -7.9 \\ \hline
2${}^{nd}$ order momentum compaction, $\alpha $${}_{1}$ & 4.68 \\ \hline
Beam pipe half-height & 50 mm \\ \hline
\end{tabular}
 \caption{Parameters of the accumulator \cite{Pos-12}}\label{tab:accu}
\end{table}


\subsection{Beam distribution onto the horn system}

\begin{figure}[htbp]
\begin{center}
\includegraphics*[width=6.10in, height=1.32in, keepaspectratio=false]{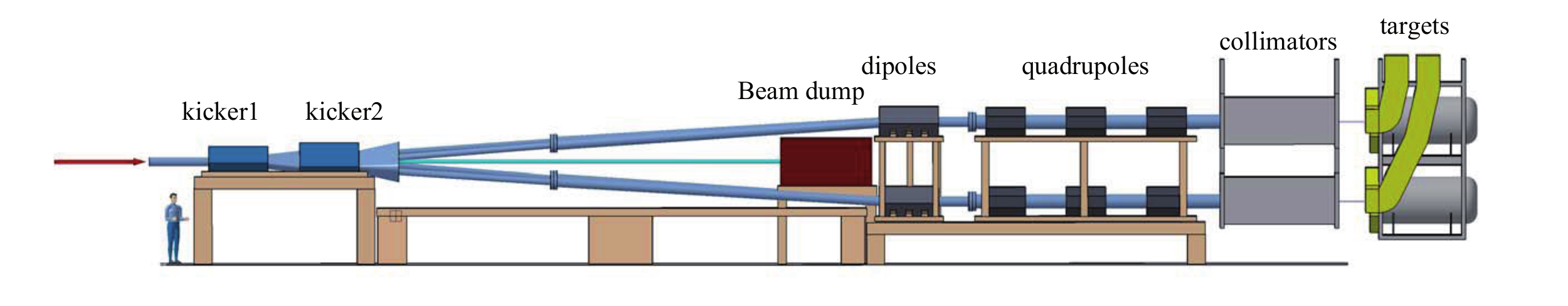}
\caption{Side view of the distribution system. } \label{fig:config3good}
\end{center}
\end{figure}

  The incoming proton beam from the accumulator needs to be split 
into four different beams and impinged on the four target-horn system at a frequency of 12.5 Hz. The general conceptual layout of the beam distribution is presented in Figure~\ref{fig:config3good}.

  The four targets are separated by a distance of 2000 mm (centre-to-centre). This value is a key parameter in the design of the beam distribution system as it determines the angle of deflection and thus the magnetic field mandatory for the splitting of the proton beam. The requirement on the Gaussian width of the beam is 4 mm. The primary proton beam coming from the accumulator is assumed to propagate along the z axis centered onto the 4-targets-horn system; two angles of deflection are therefore needed to bring the protons to the axis of each target. The use of two bipolar kickers would then be suitable to perform this task.


The two bipolar kickers make an angle of 45 degrees with respect to the central beam axis. This rotation already introduces a first angle of deflection. Therefore, according to the polarity of the magnetic field of K1 (K2), the proton beam is distributed diagonally to the compensating dipoles D1 or D3 (D2 or D4) which deviate the beam to the corresponding target T1 or T3 (T2 or T4). To avoid unwanted optical phenomena such as achromaticity and to have the beam hitting the target with an incident angle of 90 degrees the optical system has to be symmetrical. Therefore compensating dipoles (D1, D2, D3 and D4) are needed.


  A diagram of the operation mode of each optical element involved in this configuration can be therefore suggested.
A few ms before the protons enter the kicker system, the magnetic field of one of the two kickers increases to reach its maximum value. When getting between the magnets of the kicker, the protons are subject to the magnetic force induced and then are deflected by the angle $\pm$ $\alpha$ to the corresponding compensating dipole. The repetition rate for the whole horn system is 20 ms (50 Hz) which gives a rate of 80 ms (12.5 Hz) for each of the targets. Having two kickers in series implies the use of a consequent aperture of the second kicker in order to prevent the beam kicked from the first kicker to hit the magnet of the second one.

 At a distance of 15 m and at a proton energy of 4.5 GeV, the kickers must induce a magnetic field of 0.86 T to deflect the beam to the axis of the compensating dipoles. The vertical aperture of the second kicker (K2) should be at least 60 cm to allow the beam to pass through without damaging the magnets. 
The code TRANSPORT \cite{Roh-71} was used to estimate the size of the beam envelope between the kicker and the compensating dipole travelling through the four beam lines. The aim of the four beam lines is not only to distribute the proton beam to the horn system at a frequency of 12.5 Hz but also to deliver a beam having the optimum characteristics mandatory in the process of generating pions. The beam waist must be located in the middle of each of the targets (which are 78 cm long) and must have a regular Gaussian shape of width $\sigma $= 4 mm.

 A beam dump will be located after the pair of bipolar kickers in order to stop the 4.5 GeV energy proton beam in case of failure of the magnets.  For a single pulse (1.1x10${}^{14}$ protons) the power of the beam to be stopped represents 80 kW.

\subsection{Beam focusing }

  To efficiently focus the beam onto the horn system the use of optical elements such as quadrupoles is mandatory.
 Several configurations have been investigated with the code TRANSPORT including two or three quadrupoles.




  
The transverse size and the emittances of the proton beam entering the switchyard are considered to be similar to those of the beam leaving the SPL: $\sigma $ = 2 mm and the rms emittances $\varepsilon $${}_{x}$ = $\varepsilon $${}_{y}$ = 3 $\pi $ mm.mrad (Gaussian) (Table \ref{tab:accu}). The relative errors on the emittances are considered to be 20\%  
and are included in the simulations. A 1m drift is considered between the entrance of the switchyard and the location of the first kicker. This is to foresee any eventual beam monitoring at this place to check the characteristics of the proton beam coming from the accumulator.

\begin{figure}[htbp]
\begin{center}
\includegraphics*[width=5.34in, height=3.17in, keepaspectratio=false]
{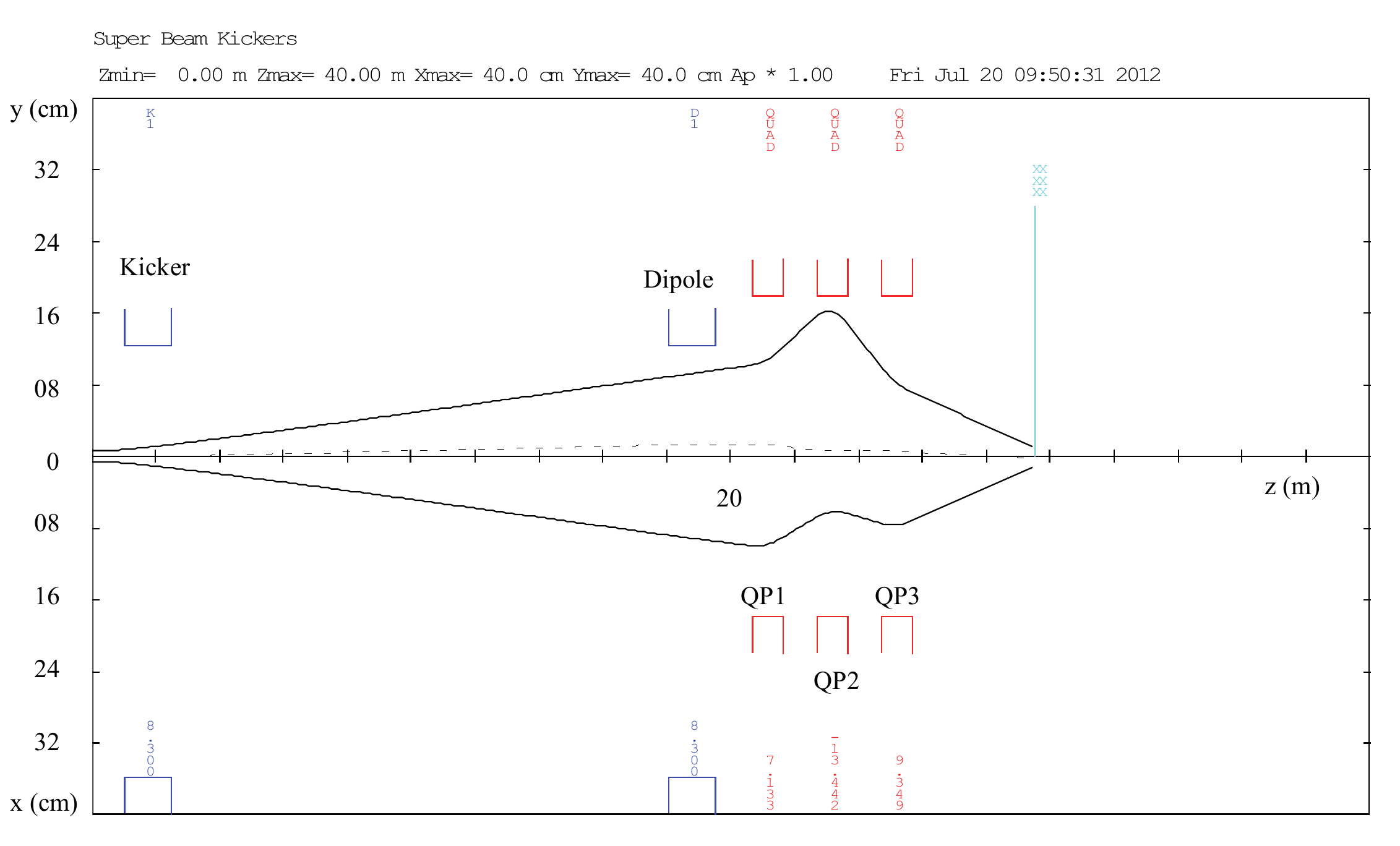}
\caption{Transverse beam envelopes. }\label{fig:config3}
\end{center}
\end{figure}

The baseline configuration is \textbf{K-D-QP-QP-QP-T}. The three quadrupoles are here placed after the compensating dipole (Figure \ref{fig:config3}).

  The beam focuses at 29.9 m (total length of the beam line and middle of the target) and its dimensions (3$\sigma $) reaches closely the values needed at this point.

  The distance between:

-the 1st kicker and the target station is 29.9 m

-the 1st kicker and the dipole 1 and dipole 3 is 17 m

-the 2nd kicker and the dipole 2 and dipole 4 is 14.7 m

-the dipoles 1,2,3,4 and the target station is 11.9 m
  A transfer line will be present between the accumulator and the transport system described here. The distance from the exit of this transfer line to the entrance of the 1${}^{st}$ kicker is assumed to be 1 m. This length is not a fixed value yet as it strongly depends on the diagnosis devices needed to control the quality and the position of the proton beam once it leaves the accumulator. 

  Table \ref{tab:summarypara} summarizes the physical parameters calculated for the kickers, dipoles and quadrupoles for each beam line of the switchyard system. 

According to the high values of the intensity needed for the coils, the use of superconducting magnets can be considered here and will be investigated in further studies.

\begin{table}[htbp]
 \centering
\begin{tabular}{|p{1.85in}|p{0.55in}|p{0.7in}|p{0.55in}|p{0.7in}|} \hline
\textbf{} & \textbf{Kicker1} & \textbf{Dipole1,3} & \textbf{Kicker2} & \textbf{Dipole2,4} \\ \hline
Field strength (T) & 0.83 & 0.83 & 0.96 & 0.96 \\ \hline
Angle of deflection (mrad) & $\pm$83.0 & - & $\pm$96.0 & - \\ \hline
Magnetic length (m) & 1.5 & 1.5 & 1.5 & 1.5 \\ \hline
Aperture H/V (mm/mm) & 250/350 & 250/250 & 250/600 & 250/250 \\ \hline
Total intensity (kA) & 115.6 & 82.6 & 152.6 & 95.4 \\ \hline
\end{tabular}
\linebreak

\begin{tabular}{|p{1.5in}|p{1.0in}|p{1.0in}|p{1.0in}|} \hline
\textbf{} & \textbf{Quadrupole1} & \textbf{Quadrupole2} & \textbf{Quadrupole3} \\ \hline
Field gradient (T/m) & 0.71 & 1.34 & 0.93 \\ \hline
Aperture radius (mm) & 180 & 180 & 180 \\ \hline
Magnetic length (m) & 1 & 1 & 1 \\ \hline
Function & F & D & F \\ \hline
Total intensity (kA) & 20.3 & 38.4 & 26.6 \\ \hline
\end{tabular}
\caption{Summary of the physical parameters of:
kicker 1,2 and dipole 1,2,3,4 (top)
quadrupole 1,2,3 (bottom) of the 4 beam lines.}\label{tab:summarypara}
\end{table}

\subsection{ Additional beam instrumentations}

  During the experiment the quality and the position of the beam must be controlled at several positions along the beam lines and mainly at the entrance and the exit point of the switchyard system. Beam collimation may be needed upstream of kicker 1 to cut off any eventual halo of the beam when leaving the accumulator. The exit point of the switchyard represents the interface with the target station and the last magnet. A consequent variation of the energy of the proton beam coming from the SPL-accumulator may also induce chromatic focusing errors within the system. The addition of sextupoles may be required to correct this effect. Beam monitors should also be added at the exit point of the switchyard to measure the transverse position of the beam and then avoid the beam from not hitting the centre of the targets as evoked in the previous section. To suppress any eventual halo from the beam and to cope with beam fluctuations (see previous section), one could consider a collimator at the exit point of the system.

\subsection{ The beam window}

The 50 Hz proton beam will be distributed equally between four targets, each of which will require its own beam window to separate the target station from the vacuum region of the four beam pipes. As the SPL beam is split into many low intensity pulses, the main challenge is not to withstand the per-pulse thermal stress, but to remove the heat fast enough so that the window does not melt or fail by an accumulation of thermal deformation over many pulses. 

\begin{figure}[htbp]
\begin{center}
\includegraphics*[width=9cm]{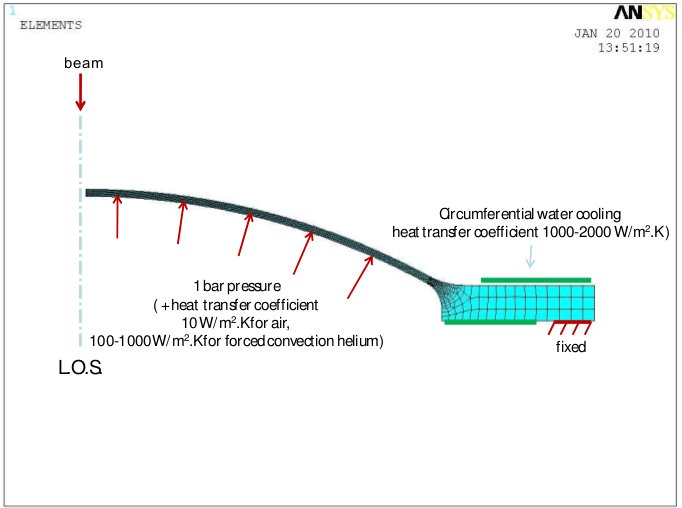}
\caption{Setup of the finite element analysis model to study the beam window. } \label{fig:bwansys}
\end{center}
\end{figure}

\begin{figure}[htbp]
\begin{tabular}{cc}
\includegraphics*[width=7cm]{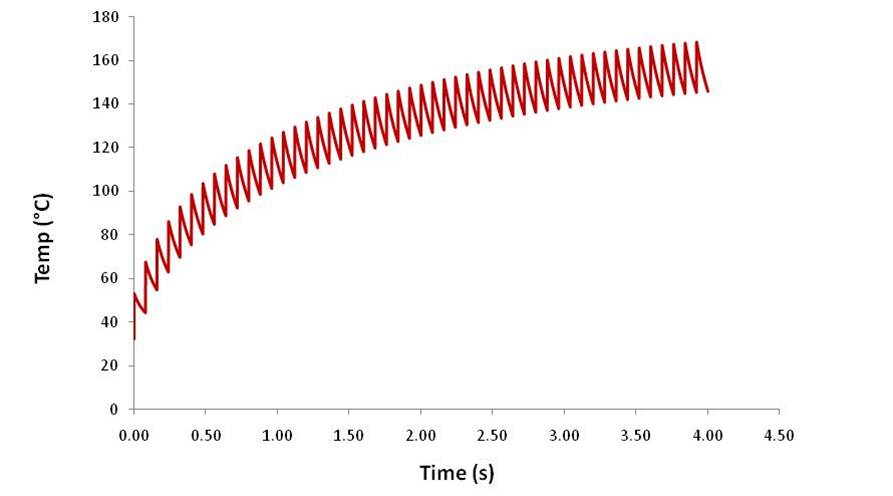} & 
\includegraphics*[width=7cm]{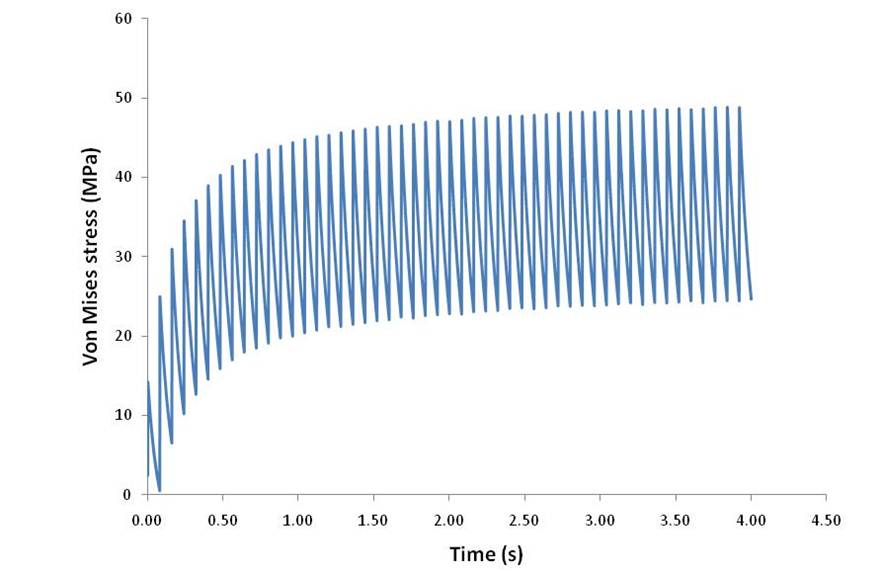} \\
\end{tabular}
\caption{Temperature (left) and von Mises stress (right) for a 0.25 mm thick beryllium window circumferentially cooled by forced convection water. } \label{fig:bwtemp}
\end{figure}

Finite element analysis studies (Fig. \ref{fig:bwansys}) 
have concluded that beryllium windows, circumferentially cooled by forced convection water cooling, are a suitable design solution. Beryllium has excellent thermal properties and simulations show that water cooling will be enough to keep the maximum temperature at beam spot region below 200 degrees Celsius. The window should be a thin (less than 0.5 mm) to reduce the beam loss and  have a partial hemisphere shape in order to withstand the differential pressure force between the target station and beam pipes.
The temperature and von Mises stress computed using ANSYS for a 0.25 mm thick beryllium window circumferentially cooled by forced convection water (assuming a heat transfer of 2000 W/m$^2$K) are shown in Fig.~ \ref{fig:bwtemp}. 

The windows should be remotely replaceable and this could be achieved using inflatable bellowed seals either side of the beryllium window. A similar design  has already been employed successfully in the Japanese T2K neutrino facility.

\section{The target station 	}
\label{sec:ts}

\subsection{Introduction}

The target station contains sets of four baffle/collimators, targets and magnetic horns within a single large helium vessel, along with the 
beam diagnostics and support infrastructure necessary for the safe and reliable operation of these components. 
The target station is separated from the primary beamline at the upstream end by four beam windows through which enters the split proton beam. 
The four split proton beams pass through the collimators, targets, magnetic horns and decay volume, before being absorbed by the beam dump/hadron absorber. 

The design of the target station must meet a number of important technical challenges. It requires substantial steel and concrete shielding. Due to the harsh conditions, horn and target lifetimes will be limited; multiple failures are expected during the lifetime of the facility. 
It is therefore essential that broken parts can be replaced, and due to the high activation this must be done using a remote handling system. It is also important that the horns and targets can be aligned with the incoming proton beams to sub millimetre accuracy. 
The use of four parallel horns will introduce further challenges unique to this facility. 
The cross section area of the beamline is increased by a factor of four, so a much larger volume of radiation shielding will be required to surround it. 
Having to accommodate four horns will increase the complexity of many operations, such as supporting the horns, connecting cooling and other services, replacing broken horns, and disposing of activated components.
 
The main objectives of the design process were as follows:
\begin{itemize}
\item	To ensure safe operation, and compliance with all applicable radiation limits. This includes ensuring the safety of repair workers and planning for the safe disposal of radioactive scrap.
\item	To minimise the amount of downtime required for repairs and maintenance. This will involve increasing the reliability of components and decreasing the time taken for repairs.
\item	To minimise the cost of construction, operation and maintenance over the lifetime of the facility. The proposed design aims to deliver a compromise between reducing cost and reducing downtime.
\end{itemize}

The starting point for the design was the target station for the T2K experiment \cite{ref:t2ktarget}, located at the J-PARC facility in Japan. 
The T2K target station was designed to allow up to 4MW beam power, and has a remote handling system with similar capabilities to those required here. 
This is a proven design which has been running for 3 years (total 3x$10^{20}$ 
protons on the first target), and  is a valuable source of practical experience. Particular attention was paid in this new design to (i) reducing the time required to change a target or horn, and (ii) reducing the generation of tritium from the concrete within the helium volume.

\subsection{Design Overview}

The requirement for remote handling will be met by using an overhead gantry crane to insert and remove components from the beamline. All four horns will be mounted on a single support module which will provide support and alignment, and allow the horns to be lifted by the crane. The horns will then be moved to a maintenance area away from the beam for repair and disposal. This maintenance area will consist of the hot cell, where human operators can carry out repairs using remote manipulators, and the morgue, where activated scrap can be safely stored.
In order to gain access to the components, the radiation shielding above them must first be removed. This will be achieved by making the top layer of shielding from movable concrete blocks which can be lifted off by the gantry crane. The beamline and maintenance area will be located at the bottom of a 10m deep pit in order to prevent radiation shine to the outside when moving components.
The target station vessel will be filled with helium at atmospheric pressure, in order to minimise pion absorption, tritium and NOx production, and thus to provide an inert environment for the target and horns. The helium will be contained in a steel pressure vessel which will surround the horns, targets, collimators and beam dump. Beam windows will be required to separate the helium environment from the accelerator vacuum. The helium vessel will have a removable lid to allow access to the components inside. 

In addition to the beamline and maintenance area, the target station must also contain the following systems;
\begin{itemize}
\item	Cooling plant for the beamline components
\item	Power supply for the magnetic horns
\item	Air conditioning system for the buildings
\item	Pumps to fill and empty the helium vessel
\item	Control room for the crane and other target station systems
\end{itemize}

\begin{figure}[ht]
\begin{center}
\includegraphics[width=0.85\columnwidth]{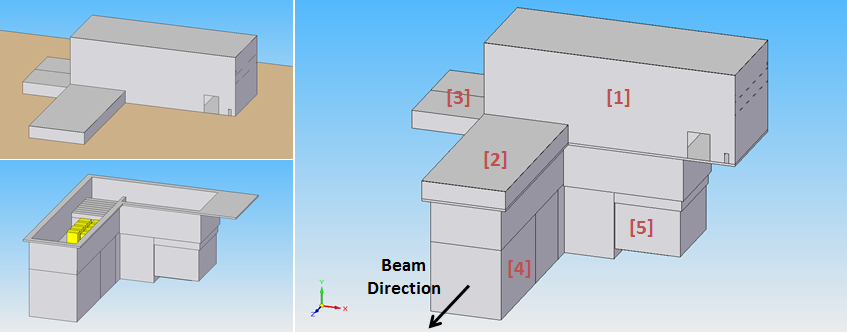}
\caption{Site Layout; 1) Main Hall, 2) Side Hall, 3) Pump House, 4) Beamline Shielding, 5) Maintenance Area}
\label{FigTS:fig01}
\end{center}
\end{figure}

The proposed layout will consist of three buildings; a main hall containing the crane, maintenance area, and access to the beamline, a side hall containing the horn power supplies and beam dump, and a pump house for cooling and air conditioning systems. In addition to the surface structures there will be a large underground volume beneath the main hall and side hall. This will contain the beamline and maintenance area, plus shielding. The overall layout of the site is shown in figure \ref{FigTS:fig01}.

\subsection{Helium Vessel}

\begin{figure}[ht]
\begin{center}
\includegraphics[width=0.85\columnwidth]{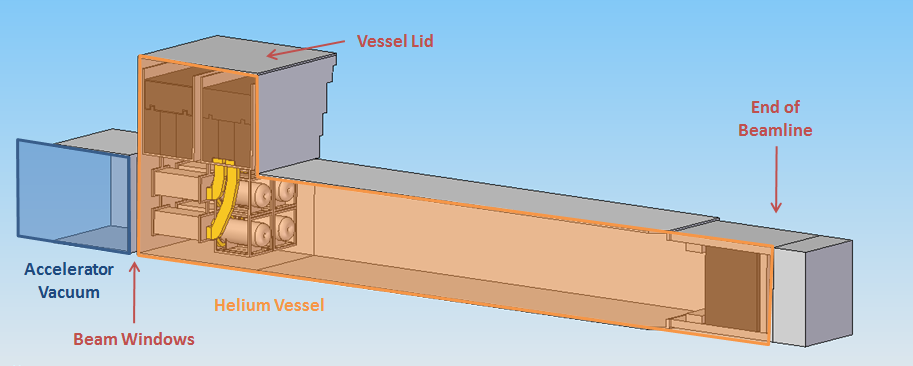}
\caption{Extent of Helium Vessel}
\label{FigTS:fig02}
\end{center}
\end{figure}

The horns, targets, collimators, decay volume and beam dump will be contained in a steel vessel filled with helium at atmospheric pressure. Figure \ref{FigTS:fig02} shows the area covered by this vessel. Using helium will avoid the problems caused by passing a proton beam through air, such as the production of nitric acid which causes steel embrittlement, and the activation of large volumes of air. Using helium rather than a vacuum will allow for cooling of components by conduction and convection, and will prevent stresses in the vessel due to differential pressure. However, the vessel will be required to temporarily withstand vacuum pressure as it will be vacuum pumped and then back-filled with helium in order to achieve a high helium purity. 
The split proton beam will enter via four beam windows which will separate the helium vessel from the accelerator tunnel. The beam windows will connect to both sides via inflatable pillow seals, as used in T2K \cite{ref:t2ktarget}. The benefit of pillow seals is that they can be remotely disconnected and do not depend on a mechanism to operate, so a damaged beam window can be replaced without requiring complex tooling or exposing a human repair worker to radiation. All four beam windows will be mounted on a single frame which can be lifted out by the gantry crane after the pillow seals have been disconnected. The hot cell will then be used to replace the damaged window without having to scrap the whole frame.

The helium vessel and the decay volume will be joined to form a single pressure vessel, as in T2K. As a result, the entire decay volume must be pumped out every time the helium vessel needs to be opened in order to replace a component. 
Running the four targets at full power will cause a predicted heat load of 511kW on the walls of the helium vessel and decay volume due to secondary particle interactions. As a result the walls will require active cooling, which will be achieved by using water channels on the outside of the vessel as for T2K.

\subsection{Support Module}

\begin{figure}[ht]
\begin{center}
\includegraphics[width=0.85\columnwidth]{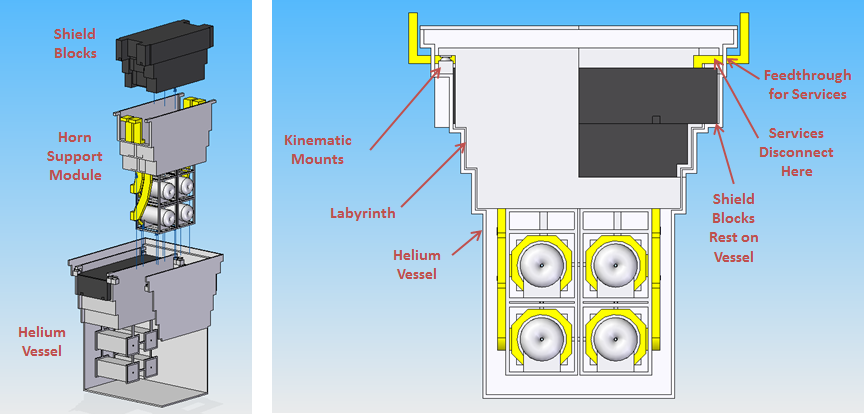}
\caption{Support Modules, Shield Blocks and Helium Vessel}
\label{FigTS:fig03}
\end{center}
\end{figure}

The horns and collimators will be held in place by support modules which can be lowered vertically into the helium vessel by crane, as shown in figure \ref{FigTS:fig03}. One support module will hold the four horns, and a second will hold the four collimators. The support modules rest on kinematic mounts at the top of the helium vessel. Removable shield blocks will fit inside the support modules, and rest on the sides of the vessel. The sides of the shield blocks will be stepped to create a labyrinth, preventing direct shine of radiation to the top of the vessel.
 The easiest place to disconnect services will be immediately after the feedthroughs, just inside the vessel. This will allow the connection points to be accessed from the top of the vessel without having to remove the shield blocks first.  A mechanism for quickly disconnecting striplines has been developed by Fermilab, and a similar design could be used here.

\subsection{Horn Alignment}

It is essential that the four horns containing the four targets can be aligned with the four proton beams to sub millimetre accuracy. This will depend on the alignment of the horns relative to the support module carrying them, and the alignment of the module relative to the helium vessel. This will be achieved by having the support modules rest on kinematic mounts, which are designed to exactly constrain the six degrees of freedom of motion. The kinematic mounts will allow the position of the module to be precisely defined in relation to the helium vessel, with high repeatability. 

\subsection{Horn Power Supply}

Power for the magnetic horns will be provided by 8 power supply units (PSUs) connected to the horns by striplines. The length of stripline required should be minimised in order to minimise electrical losses. However, the PSUs must be located outside the radiation shielding to protect them from damage. This will be achieved by locating the PSUs on top of the decay volume shielding. This ensures the PSUs are as close as possible to the horns. Above the beam dump shielding there will be space available for broken PSUs to be moved for maintenance. A 5 tonne gantry crane in the side hall will be used to carry the PSUs.

\begin{figure}[ht]
\begin{center}
\includegraphics[width=0.85\columnwidth]{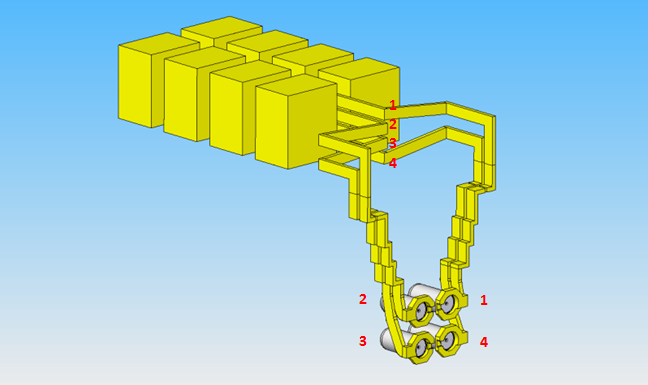}
\caption{Stripline Connections from PSU to Magnetic Horns}
\label{FigTS:fig05}
\end{center}
\end{figure}

The power supply is designed so that every horn must be connected to every PSU. The length of stripline must be roughly the same to each horn in order to ensure accurate timing, which is made more difficult by the fact that the lower horns will be further from the supply. Figure \ref{FigTS:fig05} shows the stripline layout which was designed to solve this. The length of stripline between the end of the PSUs and each horn is 20m, which is less than the specified maximum length. Each horn must be powered in turn as the beam is cycled around the four targets. Figure \ref{FigTS:fig05} shows which stripline connects to each horn, and also indicates the order in which the horns will be powered.   

\subsection{Hot Cell}

The hot cell will consist of a safe containment area for activated components and a shielded operator room. Repairs can be carried out by a human operator using remote manipulators to safely work on highly activated components. Two lead glass windows positioned at 90° to each other provide direct visibility. The crane could be used to lift and rotate the component by 180°, in order to give a complete 360° view. Access to the hot cell will be via a shaft from the control room building, allowing the hot cell to be accessed without having to enter the main hall. The roof of the hot cell will consist of removable concrete shield blocks, so it can be sealed when not in use. The roof of the operator room will also be made from shield blocks to allow for easy installation of manipulator arms using the main crane. 

\subsection{Morgue}
 
The morgue will consist of a large underground space in which broken parts can be stored until their activation level has dropped enough that they can be moved elsewhere. Components in the morgue will be sealed in steel casks to stop most of the radiation. In addition to the casks, the morgue will be shielded by concrete walls on all sides. The morgue size specified here will have enough space to contain 6 complete support module assemblies, although one of these spaces will be filled by the spare horn assembly. 

\subsection{Shielding}

\begin{figure}[ht]
\begin{center}
\includegraphics[width=0.85\columnwidth]{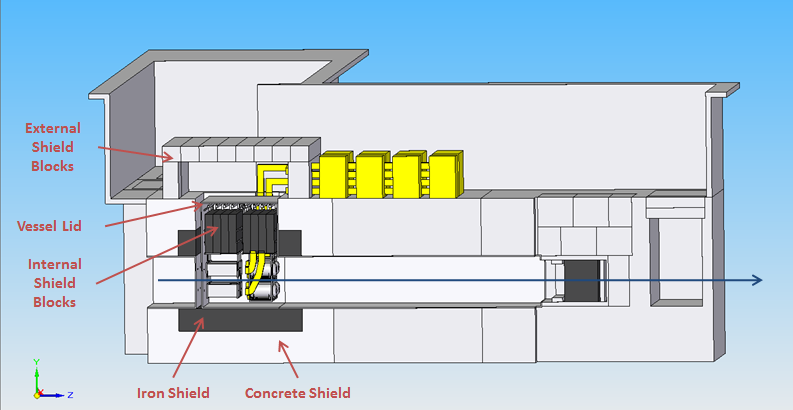}
\caption{Beamline Shielding}
\label{FigTS:fig07}
\end{center}
\end{figure}

The main source of radiation will be the horns and targets. To provide a biological shield, the helium vessel will be surrounded on all sides by a 2.2m thick iron inner shield followed by a 3m thick concrete outer shield. The rest of the beamline will be less active, and will be surrounded by a 5.2m thick concrete shield. The maintenance area will also require shielding, around 2m of concrete on all sides. Based on previous experience, it is recommended that low sodium concrete be used for the shielding, to limit the formation of radioactive sodium isotopes in the shielding. The outer concrete shield will need to be sealed to prevent activated air leaking from the region immediately surrounding the helium vessel into the target station atmosphere.
There must be some way to open the shielding in order to gain access to the components inside. This will be achieved by making the top of each shield out of stacked concrete blocks which can be moved by the crane. Figure \ref{FigTS:fig07} shows the shielding arrangement around the beamline.

\begin{figure}[ht]
\begin{center}
\includegraphics[width=0.85\columnwidth]{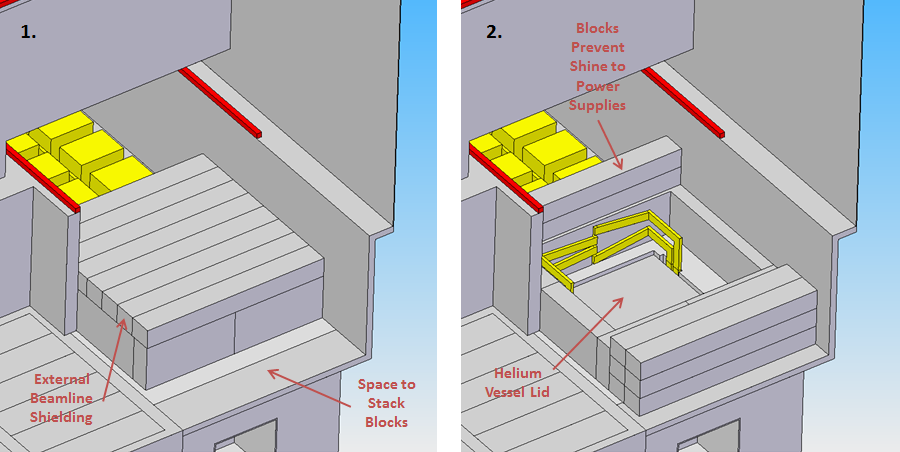}
\caption{External Shielding of Helium Vessel, 1) Closed and 2) Open to Access Vessel}\label{FigTS:fig08}
\end{center}
\end{figure}

\subsection{Crane and Control Room}

Activated components will be moved using a 100 tonne gantry crane. This crane will also be used for initial installation of components and for moving replacement parts into the target station. There will also be a 5 tonne gantry crane in the side hall, for carrying power supply units. The 5 tonne and 100 tonne cranes will overlap, so that the PSUs can be delivered to the main hall, unloaded by the large crane, then transferred to the small crane to be installed in the side hall. 


\subsection{Maintenance Procedure}

To minimise downtime, two assemblies of four horns each will be used at any one time. This will allow one assembly to be repaired while the other is running, so the beam will only have to be stopped for long enough to exchange the assemblies. The spare assembly will still be fairly active, and will therefore be stored in the morgue for safety. A procedure for a standard repair operation, for example replacing a broken target, has been studied.

\subsection{Decay Volume}

The decay volume will consist of a 25m long steel pressure vessel connecting the target station helium vessel to the beam dump. It will be directly connected to the helium vessel and so will also be filled with atmospheric pressure helium. The entire vessel will be built to withstand a vacuum when the helium is pumped out. The decay volume will be shielded with 5.2m thick concrete on all sides. The steel vessel will experience significant heating from particle interactions and will therefore require its own cooling system.
Downstream of the beam dump will be a pit to house muon monitors if required. 

\subsection{Beam Dump}

\begin{figure}[htbp]
\begin{tabular}{cc}
\includegraphics*[width=7cm]{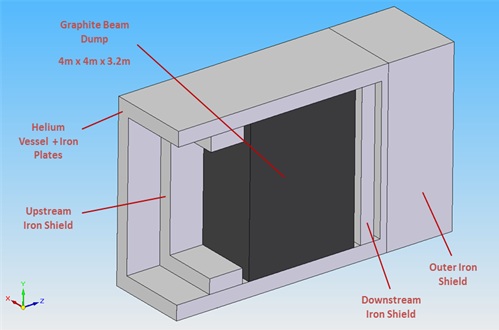} & 
\includegraphics*[width=7cm]{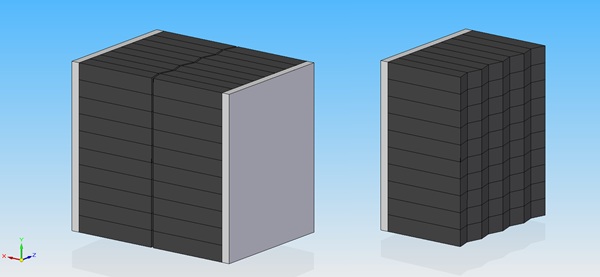} \\
\end{tabular}
\caption{Components of the beam dump (left) and graphite blocks (right). } \label{fig:bdTS1}
\end{figure}

The beam dump will consist of  graphite blocks, water cooled on two faces. The incoming proton beam does not interact directly with the cooling water, in order to prevent water hammer and cavitation. To prevent graphite oxidation, the beam dump will be contained in the same helium vessel as the target station and decay volume. The graphite will be surrounded by iron plates, to reduce radiation dose to the surroundings. An upstream iron shield is designed to act as a collimator which will protect the cooling and diagnostic systems around the beam dump. Figure~\ref{fig:bdTS1} shows the components of the beam dump.
 
It is not possible to manufacture a single piece of graphite of the required size, so the beam dump must be built up from smaller blocks  (Fig.~\ref{fig:bdTS1}). The end of each block will be cut at an angle, to prevent a direct shine path through the centre of the beam dump. The proposed grade of graphite to be used is Sec Carbon Ltd PSG-324, the same grade which was used for the T2K beam dump.

\begin{figure}[htbp]
\begin{tabular}{cc}
\includegraphics*[width=7cm]{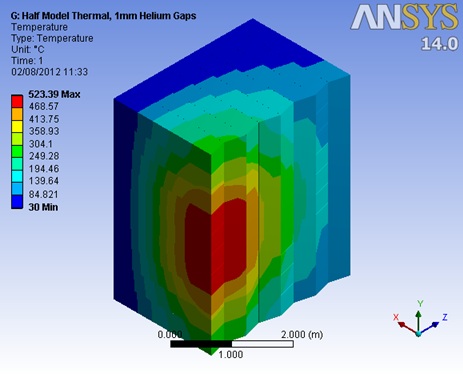} & 
\includegraphics*[width=7cm]{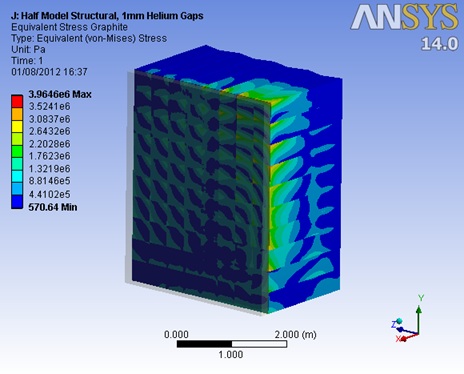} \\
\end{tabular}
\caption{Temperature  (left) and von Mises stress (right) for the beam dump. } \label{fig:bd2}
\end{figure}

Simulations were carried out in ANSYS to determine whether the proposed materials and design would be suitable. The graphite temperature was modelled based on conservative assumptions for heat transfer, with results as shown in figure 3. The body temperature results were then used as input to a structural analysis, with stress results as shown in Fig.~\ref{fig:bd2}.

The rate of graphite oxidation increases with temperature, so the required purity of the helium environment will depend on the maximum temperature. Figure~\ref{fig:bd2}  shows a maximum temperature of 523°C. Based on the limits set out by T2K this means that the required helium purity will be better than 30ppm O2, which should be feasible. The thermal performance of the proposed design should therefore be acceptable. 

Figure~\ref{fig:bd2} shows a maximum von Mises stress in the graphite of 3.96 MPa. This is fairly close to the tensile strength of the graphite (5 MPa). However, the majority of stress appears to be caused by the method of restraint rather than the actual thermal expansion. 
The proposed design should therefore be considered fit for use, as long as due consideration is given to the method of restraining the graphite.

\section{The target	}
\label{sec:target}
For the EUROnu Super Beam facility a high power target is required to generate pions to be focused by a magnetic horn. The target is expected to withstand the beam induced heating and associated stresses as well as offer reliable operation whilst exposed to intense radiation. The main technical challenges are as follows:
\begin{enumerate}
\item  Heat removal. A significant heat load is deposited by the beam on the target and has to be removed reliably by the cooling system.
\item  Static and dynamic stresses. The target must withstand thermal-mechanical stresses arising from the beam induced heating of the target. 
\item  Radiation damage. Degradation of the material properties due to radiation damage must be accommodated.
\item  Geometrical constraints. The target has to fit inside the bore of the magnetic horn whilst having an appropriate geometry for effective pion production.
\item  Remote replacement. Once activated the target has to be remotely manipulated in the event of failure.
\item  Minimum expected lifetime. The target is expected operate without intervention between scheduled maintenance shutdowns.
\item  Safe operation. The target design should minimise any hazard to the personnel or the environment

\end{enumerate}

In the proposed concept, the target stands alone from the magnetic horn, has its own cooling system and can be removed and replaced remotely. A combined target and horn design has also been considered but was rejected in favour of a separate target and horn system. The reasons for this decision are discussed in section \ref{subsec:tgt:design}. Several target technologies have been considered and the most favourable concept is presented in the following sections.

\subsection{Design philosophy}
\label{subsec:tgt:design}

Two outline target design concepts have been considered, namely 
\begin{enumerate}
\item a combined target and horn inner conductor, 
\item a separate target and inner conductor, with the target supported within the horn bore.
\end{enumerate}

Studies have shown that the latter of these two options is preferable and this has been adopted as the baseline. A separate target and horn inner conductor was found to be preferable for the following reasons:

\begin{enumerate}

\item Removing the beam heating of the target and the Joule heating of the horn are both significant challenges. Separation of the two items reduces the challenge and permits separate cooling solutions. 

\item More favourable target designs and cooling options, including segmented targets are possible, since the target is not required to conduct the horn current. A segmented target offers increased tolerance to accidental off-centre beam conditions.
\item The thermal stress in the target is reduced without the additional joule heating from the horn current pulse
\item An increase in the horn inner conductor radius is possible, which significantly reduces the magnetic stress 
\item It becomes possible to tune the target and horn geometry separately, both radially and longitudinally, which permits greater scope for optimization of the neutrino yield
\item Failure modes are not combined, possibly leading to longer lifetimes for both target and horn
\item Targets can be replaced separately within the horn, reducing cost of replacement and quantity of radioactive waste.

\end{enumerate}

 \subsection{Target Cooling}

A 1~MW proton beam with a kinetic energy of 4.5 GeV deposits of the order of 50~kW of heat in a low-Z target. Both contained water and helium gas cooling have been considered.

Helium cooling is preferred because there is negligible interaction between the beam and coolant making it readily possible for the coolant to be within the beam footprint for more direct cooling of the target. Beam induced pressure waves in a gaseous coolant are largely reduced if compared with a liquid coolant, little activation of the helium is expected and there are no corrosion issues with the target and cooling circuit materials. Several different target cooling geometry options are possible. Challenges or disadvantages of helium cooling compared with water include the fact that a relatively high pressure (larger than 10 bar) is required to generate a sufficient mass flow whilst limiting gas velocity and pressure drop to acceptable levels.

\subsection{ Thermo-mechanical design of the target}

\subsubsection{Packed Bed of Titanium Spheres}

A packed bed of target spheres has been considered because of its inherent lower quasi static and inertial dynamic stresses. The packed bed target is made up of a canister containing many small target spheres. The packed bed canister has dimensions of the same order as the monolithic target but the individual spheres are much smaller. This has three advantages in terms of stress: 

\begin{enumerate}

\item The spheres are almost uniformly heated by the beam because of their size and have a much shorter conduction path from the centre of the sphere to the cooled surface. This means the temperature gradients in small spheres are small with respect to a larger monolith of the same thermal conductivity. The quasi static stresses are driven by the temperature gradient and they are correspondingly lower.
\item The expansion time of a small sphere is much shorter than that of the solid monolith of the same material. In the case of the monolith the expansion time is longer than the pulse duration and as such significant inertial stresses occur. 
With small spheres the expansion time can be less than the pulse duration and so inertial stresses as a result of rapid energy deposition are negligible.
\item In the event of an off-centre beam hitting a target an asymmetric temperature profile is set up. This will have the effect of bending a solid monolith target and producing additional stress oscillations. As the spheres in a packed bed are not connected to each other and experience a close to uniform energy deposition whether the beam is on centre or not the packed bed configuration is inherently insensitive to an off centre beam.

\end{enumerate}

Compared to the solid monolith target the packed bed has a lower density. Beryllium has been considered for the solid target which has a density of about 1.85 g/cc. The bulk density of the spheres can not exceed 74 \% of the solid density. The density of the target material has an important effect on pion yield and so in order to recover the bulk density loss, titanium which has a density of 4.5 g/cc is proposed as a candidate material. A comparable pion yield from the surface of a solid Beryllium target and a 74 \% density Titanium target has been demonstrated using detailed simulation. A Titanium packed bed has been evaluated from a physics point of view with favourable results.



\begin{figure}[ht]
\begin{center}
  \includegraphics[width=10cm]{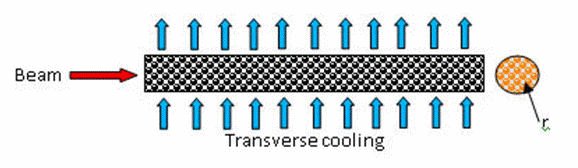}
  \caption{ \label{fig:tg6} Packed Bed ideal flow configuration. }
\end{center}
\end{figure}

The packed bed canister would have a diameter just larger than the upstream baffle to protect it from a direct hit from the beam. It would be surrounded by coolant flow channels and would be perforated to allow the coolant to pass through the centre of the spheres. This configuration gives rise to significantly more surface area for heat transfer than is present with the monolith target. The ideal flow configuration is transverse, i.e. the coolant passes through the packed bed in a direction perpendicular to the beam (Fig. \ref{fig:tg6}). This minimises pressure drop and so allows a greater volume flow through the target. As with all solid high power targets that are gas cooled an advantage can be gained by pressurising the coolant. This allows an increase in mass flow without increasing the required pressure drop to drive the gas through the target.

The larger surface to volume ratio with respect to the monolithic target and the proximity of the coolant to the core of the target offers potential for greater heat dissipation. Coolant gas is preferred over liquid due to complications associated with a liquid passing through the beam.

\subsubsection{Packed Bed Model}

\begin{figure}[htbp]
\begin{center}
  \includegraphics[width=10cm]{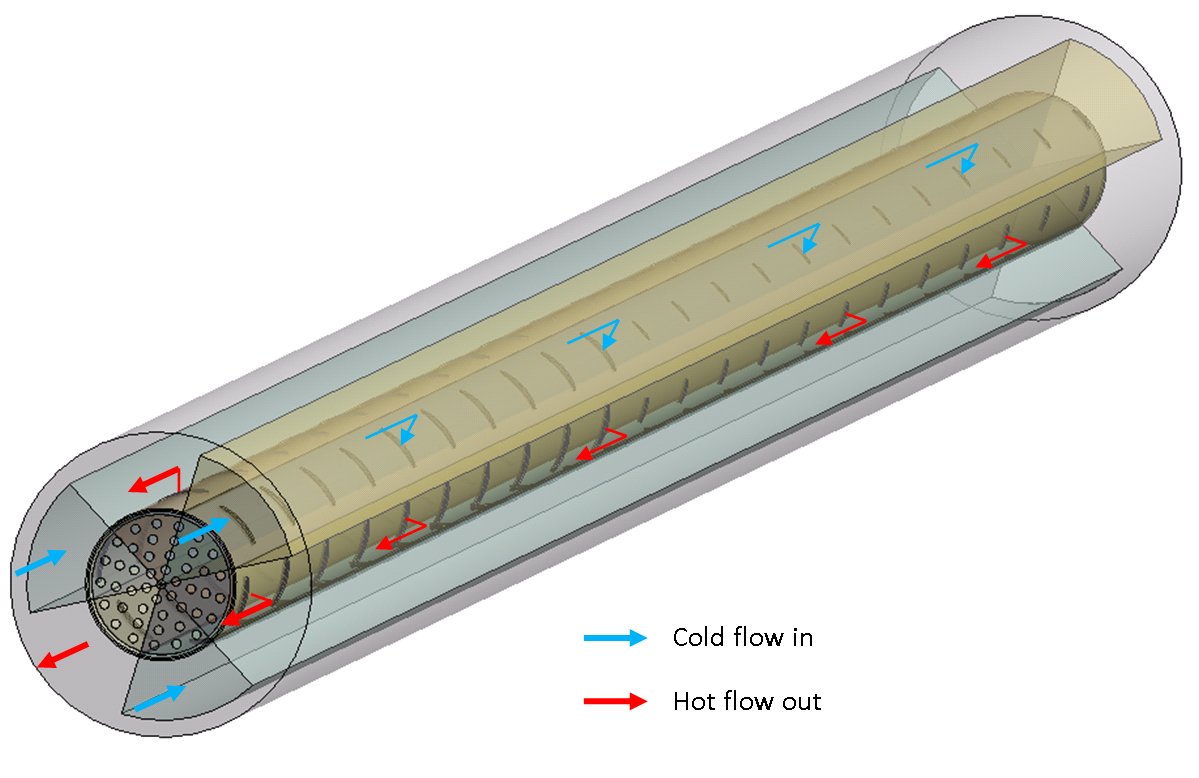}
  \caption{ \label{fig:tg6a} Proposed Packed Bed Flow. }
\end{center}
\end{figure}

An example case of a packed bed of Ti6Al4V spheres with transverse flow has been modelled. Energy deposition in the spheres has been calculated from a FLUKA model of a titanium solid target with half density. Titanium has better thermal conductivity than its alloys but some alloys such as Ti6Al4V have much higher strength and as such has been chosen for this example. Obtaining a practical transverse flow configuration within the confines of the magnetic horn is not trivial however a scheme is described here and some preliminary conjugate heat transfer modelling (using CFX) on an example case with a 1~MW beam has been carried out.  The geometry involves three inlet and three outlet channels spread at 120$^\circ$ around the canister (Fig. \ref{fig:tg6a}). Holes of various sizes are strategically placed in the canister to allow gas to flow through the packed bed. 
The packed bed is modelled as a porous domain and appears to act as a diffuser with the flow naturally dividing evenly throughout the porous domain. 
The pressure drop in the porous domain is calculated using the Ergun equation \cite{ref:target-ergun} i.e.

\begin{equation}
\Delta P=   \frac{f_p \rho V_s^2 (1-\epsilon) L}{D_p \epsilon^3},
\end{equation}

  where $f_p$ is a function of the packed bed Reynolds Number, $\rho$ is the density of fluid,
   $V_s$ is the superficial velocity, $L$ is the length of the bed 
$D_p$ is the equivalent spherical diameter of the packing,
$\epsilon$ is the void fraction of the bed.

\begin{figure}[htbp]
\begin{center}
  \includegraphics[width=10cm]{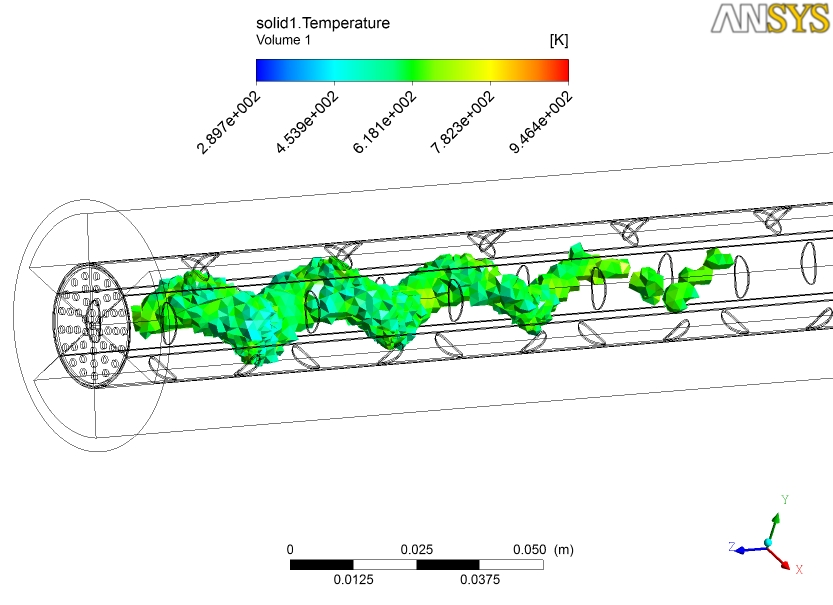}
  \caption{ \label{fig:tg11} Temperature in the packet bed target. }
\end{center}
\end{figure}

A mass flow of helium of 93~g/s is used with an outlet pressure of 10~bar. The pressure drop in the packed bed itself seems perfectly manageable and it appears as though there is scope for higher flow rates. Experience so far indicates that flow in the channels and in particular through the holes into the packed bed is the most significant cause of pressure drop. However the predicted pressure drop of 1.1 bar appears reasonable and little design optimisation has yet to be put into this example case. The maximum helium temperature is 584 $^\circ$C although the average outlet temperature is only 109 $^\circ$C.  
This difference is due to the energy deposition in the packed bed not being uniform. The maximum sphere temperature is calculated to be 673 $^\circ$C (Fig.~\ref{fig:tg11}). The maximum steady state (ignoring temperature jump) sphere temperature, $T_c$, depends on the size of the sphere, $D_p$, conductivity of the sphere material, $k$, and the surface temperature, $T_s$.$  $
\begin{equation}
T_c-T_s=\frac{Q (D_p/2)^2 }{6k}
\end{equation}
    
where $Q$ is the energy deposition (W/m$^3$). The surface temperature depends on the heat transfer coefficient between the coolant gas and the sphere. This is calculated from a Nusselt number correlation for heat transfer in pebble beds with high Reynolds number~\cite{ref:target2}
\begin{equation}
Nu= [(1.18 Re^{0.58} )^4 + (0.23 Re^{0.75}]^{0.25} 
\end{equation}

The three outlet channels are common and are configured such that the structure does not experience any significant asymmetries in its temperature profile.

Below follows a list of the key areas that need further work for the development of the packed bed target concept.

\begin{enumerate}
\item The requirement for pressurised cooling gas necessitates a beam window that can withstand the pressure difference between a vacuum and the coolant pressure.
\item Slight movements between the packed spheres may occur as a result of the sudden temperature jumps and corresponding thermal expansion. The titanium spheres with the highest energy deposition will have a temperature jump of 83 $^\circ$C with a 1~MW beam. 
\item If the beam pulse is much shorter than the expansion time of the spheres this could give rise to an additional shock stress  (assuming instantaneous heating). However the expansion time of the spheres is very small (a fraction of a microsecond) with respect to a solid monolith target so these inertial stresses are likely to be less important.  None the less this should be checked.
\item The canister would need to accommodate the thermal expansion of the target spheres.
\item The beam must pass through the canister to enter the packed bed, a perforated cooled plate is envisaged to enclose the spheres while allowing coolant to pass through so as to minimise temperature gradients in the perforated plate. Stress analysis of this component is required.
\item Off-centre beam effects on the packed bed canister should be evaluated.
\item A higher flux of neutrons is expected from the titanium packed bed with respect to the beryllium monolith. This may have a detrimental effect on the horn and needs to be investigated.
 
\end{enumerate}

\subsubsection{How much heat can be removed from a packed bed?}

The limiting factors for the heat dissipation capability of a packed bed are the coolant exhaust temperature, the coolant pressure drop across the target and the peak temperature and stress in the target spheres. 
For this 1~MW example modelled here it appears there is some head room in terms of the key limiting factors, one may even be bold enough to say that a target capable of dissipating a multi megawatt beam may be possible. This has been claimed for the case of a high Z packed bed by Sievers and Pugnat in the past \cite{ref:target3}. In order to find the practical limit of a packed bed some further analysis and Computational Fluid Dynamics is required.


\section{The horn	}
\label{sec:horn}

\subsection{Hadrons focusing system: the electromagnetic horn}

In the case of the CERN SPL Super-Beam (SB) the operation conditions of the horn will be much more severe
than in previous applications. Table~\ref{comp} shows a comparison of some horns already used 
by past or ongoing projects. In this table one can see that this horn has 
a small length which could be an advantage during the fabrication and operation, but, on the 
other side, the proton driver power (4~MW) and repetition rate (50~Hz) are considerably higher
than other applications, a real challenge!

A first step to mitigate the problem has been taken by splitting the beam onto four identical targets and horns, as described previously. In the following we study the horn for this design option.

\begin{table}[htbp]
\caption{Comparison of horns.}
\begin{center}
\begin{scriptsize}
\begin{tabular}{|c|c|c|c|c|c|c|c|}
\hline
Project             & Proton Energy & Power   & Rep. Rate       & Current  & Number of  & Length \\
                          & (GeV)                & (MW)     &   (Hz)                &    (kA)     &      horns       &   (m)     \\
\hline
\hline
CNGS              & 400                    & 0.2         & 2 pulses/6 sec & 150       & 2                             & 6.5 \\
K2K                  & 12                      & 0.0052   & 0.5                      & 250      & 2                             & 2.4--2.7 \\
NUMI               & 120                    & 0.4          & 0.5                      & 200      & 2                             & 3 \\
MiniBoone      & 8                        & 0.04        & 5                          & 170      & 1                             & 1.7 \\
T2K                  & 50                      & 0.75        & 0.3                       &   320    & 3                             & 1.4--2.5 \\
SPL-SB           & 3.5-5                 & 4              & 50                       & 300--600      & 1-2                          & 1.3 \\
\hline
\end{tabular}
\end{scriptsize}
\end{center}
\label{comp}
\end{table}%

\subsection{Horn design}

An initial design of a horn prototype system (horn+reflector)~\cite{Ball:2000pj, Gilardoni:2004kr}
foreseen for a neutrino factory (NF) has been made at CERN for a 2.2~GeV proton beam.
An optimization and a redesign has been made in a SB context ~\cite{Campagne:2004cd, Cazes:1900zz},
driven by the physics case of a long baseline experiment (130~km) between CERN and Fr\'{e}jus
(MEMPHYS detector location).

\begin{figure}[htbp]
\centering
\includegraphics[width=12cm]{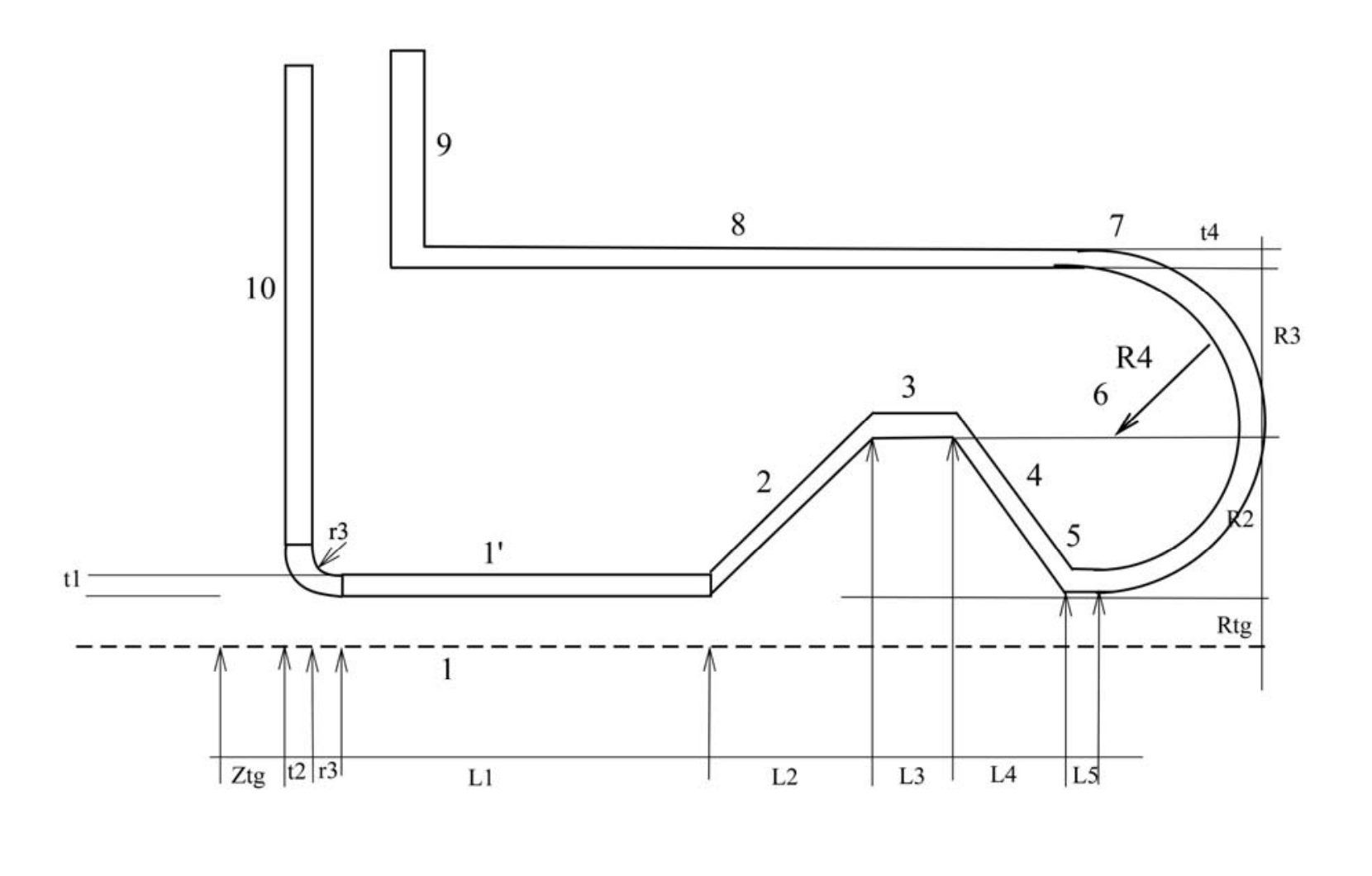} 
\caption{\label{horn_cond}Horn parameters}
\end{figure}

New studies of a hadron focusing horn have been done and as a result an optimal closed forward geometry 
with non integrated target  has been designed \cite{alonghin_19_10} ,
\cite{alonghin0},  shown in Fig. \ref{horn_cond} and with geometric parameters reported in Table \ref{tab:geometry}. 
In summary, high magnetic field closed to the target and small material thickness are  desirable to obtain the best meson
focusing and minimize multiple scattering and secondary particles interactions \cite{vanderMeer:1961sk}.

\begin{table}[htbp]
\small{
\centering
\begin{tabular}{|c|c|c|}
\hline
 Parameters &  value [mm] \\
\hline
  $L_{1}$, $L_{2}$, $L_{3}$, $L_{4}$, $L_{5}$    &  589, 468, 603, 475, 10.8 \\
\hline
$t_{1}$, $t_{2}$ , $t_{3}$, $t_{4}$ & 3, 10, 3, 10 \\
\hline
$r_{1}$, $r_{2}$ & 108 \\
\hline
$r_{3}$ & 50.8 \\
\hline
$R^{tg}$ & 12 \\
\hline
$L^{tg}$ & 780 \\
\hline
$z^{tg}$ & 68 \\
\hline
$R_{2}$, $R_{3}$  & 191, 359 \\
\hline
$R_{1}$ non integrated & 30 \\
\hline
\end{tabular}
\caption{Horn geometric parameters.}\label{tab:geometry}}
\end{table}

Given the nominal values of the proton beam power $P=4$ MW and
current $I_0=350$ kA, a high power density is present inside
the target and horn wall conductors. The feasibility of this
horn design depends mainly on the temperature and stress level
inside the target and horn structure. The stress level needs to
be compared  to the fatigue strength of the material to give an
estimate of the horn lifetime.

\begin{figure}[htbp]
\begin{center}
\includegraphics[width=8cm]{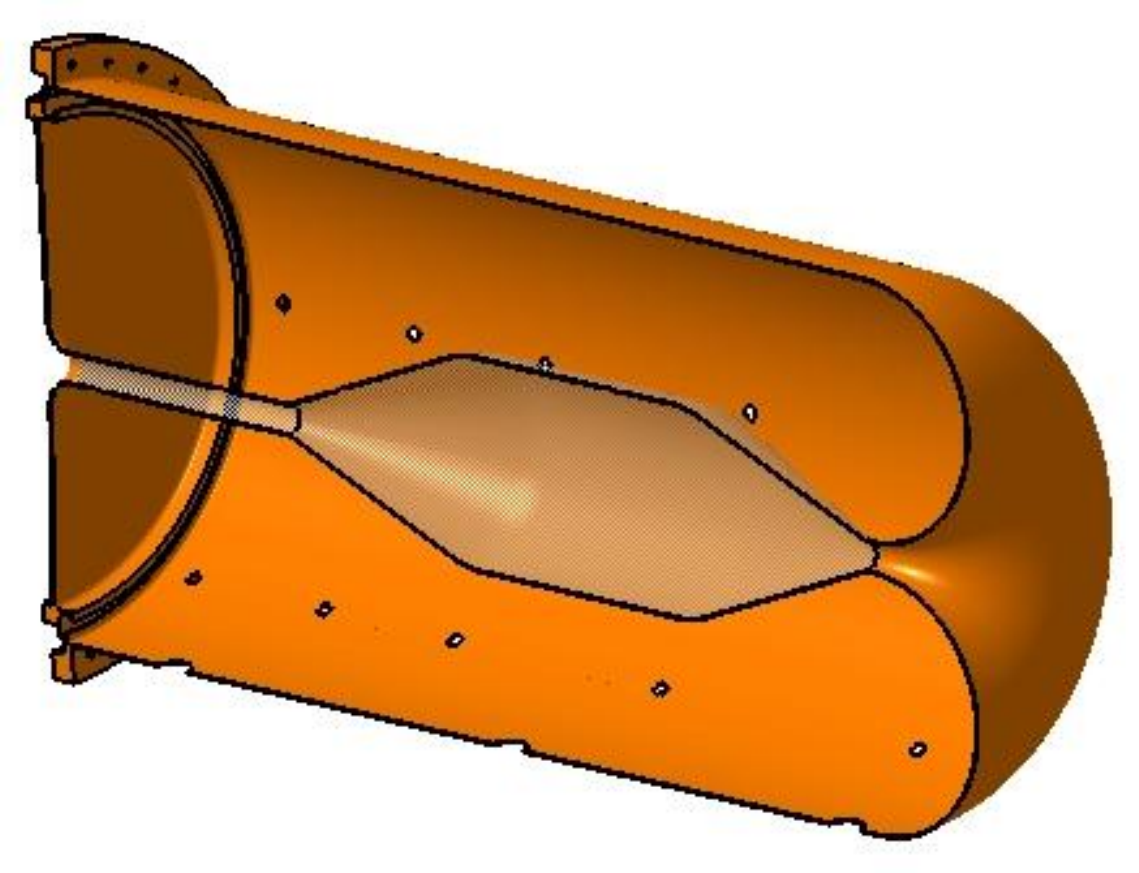}
\end{center}
\caption{\small{\label{horn} Cross section of the horn.}}
\end{figure}

The horn will be made of Aluminium AL 6061-T6 with 3 (10) 
mm thickness for the inner (outer) conductor. The horn is approximately
$2.5$ m in length and $1.2$ m in diameter. For the 
horn assembly, the different parts will be  welded at different locations, preferably
in the low stress regions. The inner and outer conductor end plates are electrically 
insulated with a glass disk  of $2.5$ cm thickness. The target with its own
cooling system will be inserted inside the central hole of the horn with  an inner
diameter of $6$ cm.  Spacers will have to be designed to maintain the target inside
 the horn.
 
In the following sections we present the electrical, thermal and mechanical studies
of the electromagnetic horn.

\subsection{Electrical currents and magnetic flux}


An analytic calculation for the toroidal magnetic field in the horn created by the alternate current has been performed. 
 Most of the current inside 
the inner conductor is 
flowing in the region $ 3.1 < r < 3.3$ cm, accordingly  to the calculated 
skin depth. Electrical losses occur in the inner
conductor, conical sections and at the top end of the horn.




\subsection{Thermal loads and cooling }



In steady state and from the power density distribution,
it is possible to calculate the required heat transfer coefficient $h$ to
 maintain a temperature difference $\Delta T = T_{horn}-T_{inf}=40\,{}^\circ C$.
The cooling efficiency of the system required to maintain a constant temperature inside the 
horn structure is proportional to the thickness wall $e$ and the power density $q$.

The temperature distribution has been computed for 
a basic cooling scenario of $\{h_{inner},h_{horn}\}=\{1,1\}$~kW/(m${}^2$K ) 
and for an optimized cooling scenario with  higher cooling in the hot spot area
$\{h_{inner}, h_{corner}, h_{conv}\}=\{3.8, 6.5, 0.1\}$~kW/(m${}^2$K).
$h_{inner}$, $h_{corner}$, $h_{conv}$ being the heat transfer coefficient 
on the surface of the inner conductor, on the upstream  bottom corner
(near the target)  and on the right side of the upstream bottom plate.

The high heat transfer coefficient seems to be quite challenging as it requires a high water flow \cite{Cooling}. Further developments are required on the basis of commercial nozzles  in order to increase the conventional capacities \cite{Analytical}. Nevertheless, heat transfer coefficient in the range of 10 ~kW/(m${}^2$K) can be expected at flow rate of approximately 4 l/min with the help of the micro-channel technique developed for VLSI chips at Soreq \cite{Highheat}.

For the uniform cooling, the maximal temperature is $180\,{}^\circ C$. When  higher cooling
is used in the hot spot area, the maximal temperature is $61\,{}^\circ C$.
The water jet nozzles disposition and  individual flow rates  of the jets 
will have to be chosen according to these $h$ coefficients required to maintain a reasonable
maximal temperature around $60\,{}^\circ C$.
This thermal model shows that the two hot areas are the upstream bottom corner 
and the downstream part where the inner radius becomes $r=3$ cm. These two 
domains will have to be cooled very well to avoid any failure.

\subsection{Static mechanical model}\label{eq:mechanics}


The displacement field has been computed and shows 
a maximal displacement of $u_{max}=1.12$ mm occurring in the downstream 
part of  the horn (opposite to the target side).

The maximal stress of $62$ MPa 
occurs in the corner region. This value is well below the aluminium maximal
strength but still high in comparison of Al 6061 T6 fatigue limit for $10^8$
cycles.  There is also a high stress level in the top inner waist of the horn. 
This part and segments junctions will require some slight modification to
achieve a stress as low as possible below $20$ MPa for example. This static
thermal stress is due to thermal gradient due to non uniform temperature 
distribution inside the horn.

It is interesting to note that the static stress level can be greatly reduced to ~6 MPa
if we achieve a uniform temperature. The displacement is about $2.4$ mm when
the horn submit to uniform thermal dilatation with $\Delta T=40{}^\circ C$.

\subsection{Transient mechanical model }

The transient stress 
from the magnetic pressure pulse is significant mainly
for the inner conductors of the horn with small radius such as the inner
conductor  parallel to the target and inner waist in the downstream
region.

\begin{figure}[htbp] 
\centering
\begin{tabular}{cc}
\includegraphics[width=8cm]{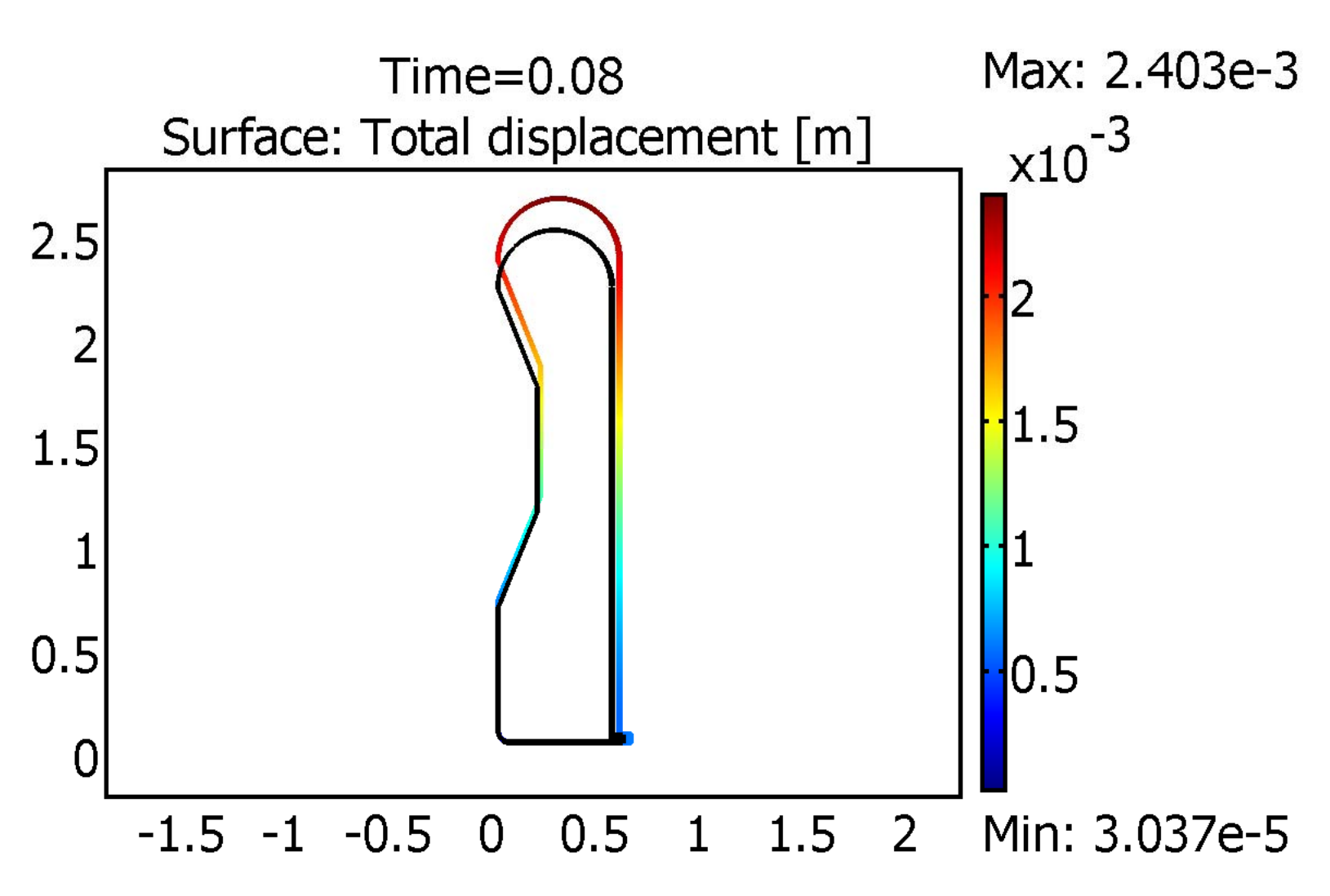} &
\includegraphics[width=8cm]{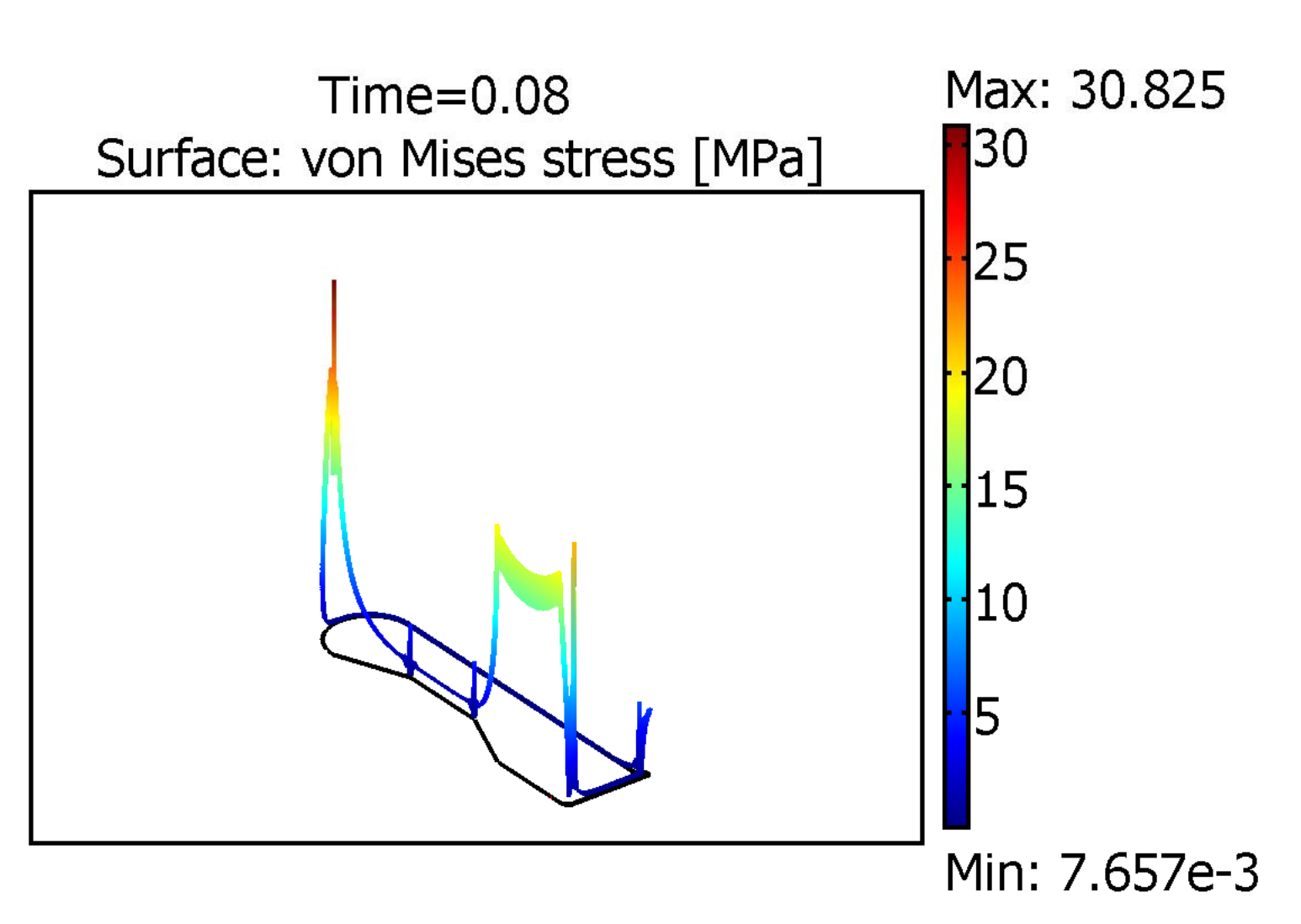} \\
a) $u_{max}=2.4$ mm, $t=80$ ms & b) Von Mises stress $s_{max}=30.$ MPa, $t=80$ ms\\
\end{tabular}
\caption{Displacement field a) and von Mises stress b) due 
to thermal dilatation with uniform temperature $T_{horn}=60 {}^\circ C $. }\label{fig:hornstress}
\end{figure}

 The displacement is maximum in the top part of the horn (downstream region, Fig.\ref{fig:hornstress}). 
The displacement due to 
the magnetic pulse is quite low in comparison to the thermal dilatation. The von
Mises stress is the highest in the upstream corner region. The magnetic pressure pulse 
contributes for about $20$ MPa in the top part of the horn region with $r=3$ cm.

 The thermal dilatation
does not contributed to the radial stress but mainly to the longitudinal stress
$S_z$ as expected. The thermal static von Mises stress is about $2.5$ MPa and the
peak stress is $15$ MPa. Because the inner conductor thickness $e=3$ mm is small
compared to the inner radius $r_i= 30$ mm the hoop stress inside the inner conductor
is approximately constant with a value of 19 MPa.

\subsection{Cooling system}

\begin{figure}[htbp]
\begin{center}
\includegraphics[width=7cm]{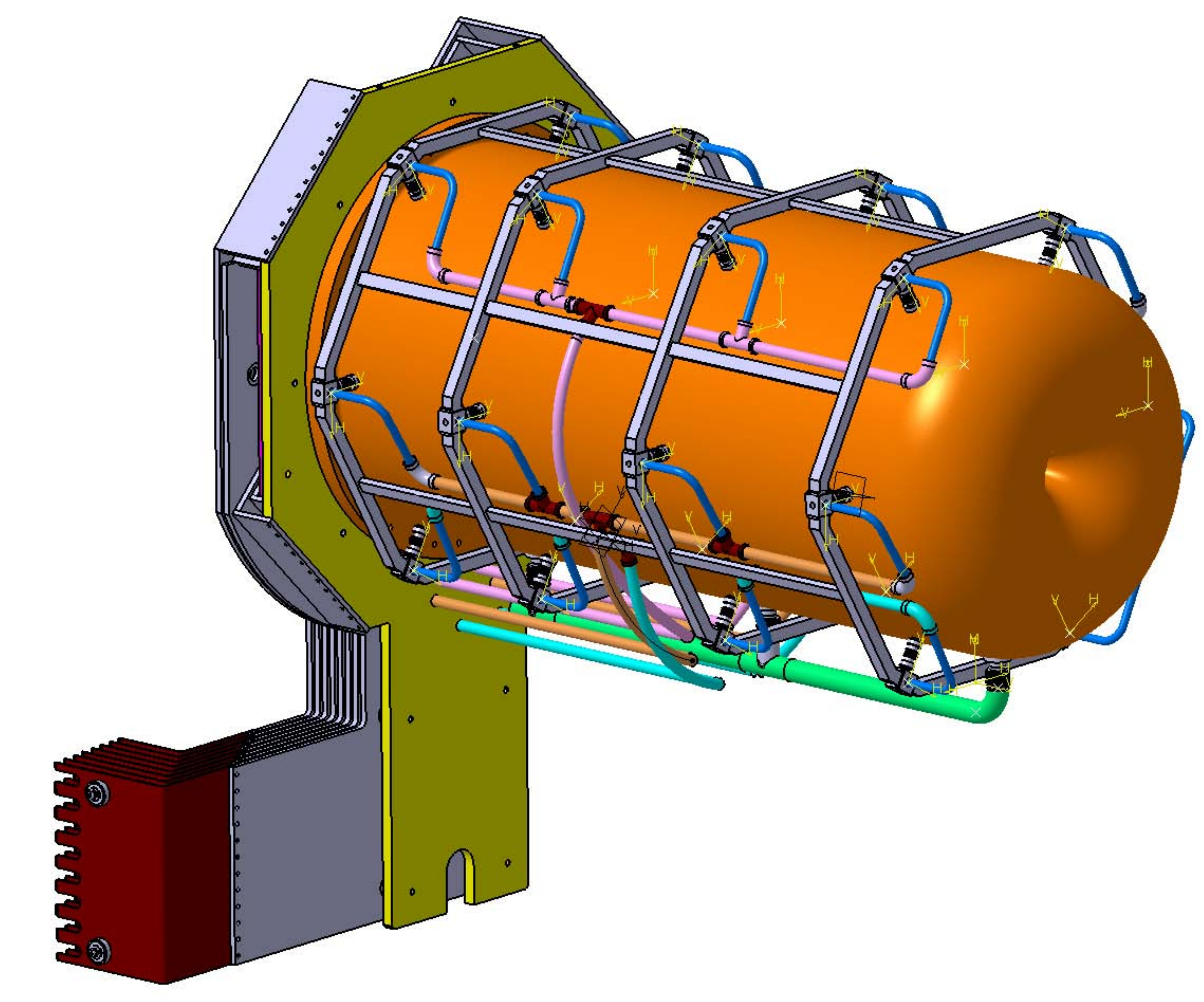}
\caption{ \label{horn_system2}Horn with striplines and cooling system.}
\end{center}
\end{figure}

The heat sources are: electrical resistive losses from pulsed currents and secondary particles generated from the
proton beam/target interaction.  
The heat transfer coefficient depends on the two water phases, the flow rate, the geometry,
and the disposition of the nozzles.
Assuming a initial inlet temperature and outlet temperature 
 $\{T_{i},T_{outlet}\}=\{20, 60\}^\circ C$ and a total power to removed of $Q=22 + 40 = 62$ kW, 
 the water mass flow rate is 0.37 kg/s. 
Hence, assuming ideal heat removal the minimum water flow rate will be 24 l/min.  The final flow rate can  be estimated to be  in the range of $60 - 120 $
l/min 
 per horn. The flow rate and jets characteristics will be chosen in order to limit the conductors temperature
below a safe limit around $60\, {}^\circ$C and to remain in a single liquid phase cooling regime.
To minimize possible failure or water leaks, it is preferable to minimize the number of jets.
Currently, 6 jets are located in the circumferential direction covering an angle of $60 ^\circ$ 
each with 5 rows giving $30$ jets in total.

\subsection{Modal analysis, natural frequency}

The current pulse circulating inside the horn is of sinusoidal form with a $100\, \mu s$ width.
The repetition frequency is $12.5$ Hz in normal use with a 4 horn system or $16.6$ Hz with 3 horns running.
The first six eigenfrequencies for this current horn geometry are $f=\{63.3, 63.7, 88.3, 138.1, 138.2, 144.2\}$ Hz excluding 
all the pipes and the frame connected to the horn outer conductor.
 The first three fundamental modes are related to 
the inner conductor vibrations, the fourth, fifth and six modes are related to the outer conductor vibrations.

\subsection{Considerations on fatigue}

The design lifetime of the horn should aim at $10^9$ pulses which is about $926$ days.
There is no fatigue limit for Aluminium alloy. Moreover the fatigue data
can only give a probability of failure for a determined level of stress
and number of cycles. In the MiniBooNE horn design
\cite{design_report_neutrino_beam} the maximum equivalent stress limit  is  $68$
MPa everywhere in the horn  to have a $97.5\%$ confidence level for no failure at $10^8$ cycles.

The presence of an initial mean stress such as mean stress due to thermal dilatation reduces the fatigue 
strength \cite{fatigue_curves}. For sustained cyclic conditions, the material should stay in the elastic regime or in other words any combination of mean stress and 
alternating stress should not create yielding or plastic deformation.

According to reference \cite{fatigue_curves}, the fatigue strength limit of dynamic stress
is  50 (20) MPa for  $10^9 $ pulses for zero (maximum) mean stress
respectively. For the weld junction with mean stress  a limit of $10$ MPa should be used.

For the inner conductor horn, the magnetic pressure pulse creates a peak
of the dynamic stress of about 16 MPa of the von Mises
equivalent stress. This value is below the 20 MPa limit strength for $10^8$  
cycles and with mean stress due to thermal dilatation  \cite{Miniboone,Fatiguel}.

\subsection{Effect of neutrons irradiation}

In the case of high neutrons flux, ($ > 6\times 10^{22} n/cm^2$) the formation of He and H creates cavities and bubbles inside the materials. These defaults lead to a reduction of the mechanical properties of the material \cite{Campagne}. Nevertheless, Fluka simulations shows that the neutrons flux through the horn is much lower than $10^{22} n/cm^2$, so the material properties should not be degraded by neutron irradiation.
The mechanical properties of the Aluminium alloy 6061-T6 may change under irradiation of all the secondary particles generated from the proton beam and target interactions and their synergy with the applied stresses \cite{Integration,Towards}. For moderate neutrons flux the neutrons create the transmutation of $Al_{27}$ to $Si_{28}$. This can lead to the formation of $Mg_{2}Si$ precipitate and an increase of the yielding strength (limit of elasticity) and the ultimate tensile strength. Radiation hardening generally decreases the tensile elongation (depending on the alloy). This issue has to be investigated in order to evaluate the impact on the material resilience in the case of fatigue stress. 


A first evaluation of the influence of irradiation on the lifetime of the
horn inner conductor indicates strong dependence of the number of cycles
to failure on the maximum dpa \cite{Fatiguel,Degrada}. A parametric study involving
both irradiation induced micro-damage (Frenkel pairs, micro-voids,
micro-cavities containing He) and mechanically induced damage fields
(micro-cracks and micro-voids) shows that for the maximum dpa not
exceeding 10$^{-5}$, the number of cycles to failure reaches more than 10$^5$.
Each higher level of dpa leads to strong reduction of the number of cycles
to failure, following a power law. In particular, a dpa level of 0.1 may
already compromise the integrity of the inner conductor. In order to
confirm these values and establish the range of safe performance of the
horn, further R\&D study is necessary. One of the crucial parameters that
still remains to be confirmed is the evolution of dpa as a function of
time (number of cycles) in the inner conductor part located in direct
proximity of the target. Such a study will result in final confirmation of
the lifetime of target-horn.



Although multi-physics simulation of the whole system can greatly help the conception of
a reliable design, a dedicated R\&D and testing with a target will be needed in the future 
to validate these studies but also to face the various safety aspects (chemistry of heavy 
metals, high radiation levels, high voltage,high current\ldots), which would also include
the design of a complete remote handling installation for the horn and target maintenance
and possible exchange.

\subsection{The horn power supply}

\begin{figure}[htbp]
\begin{center}
\includegraphics[width=11.cm]{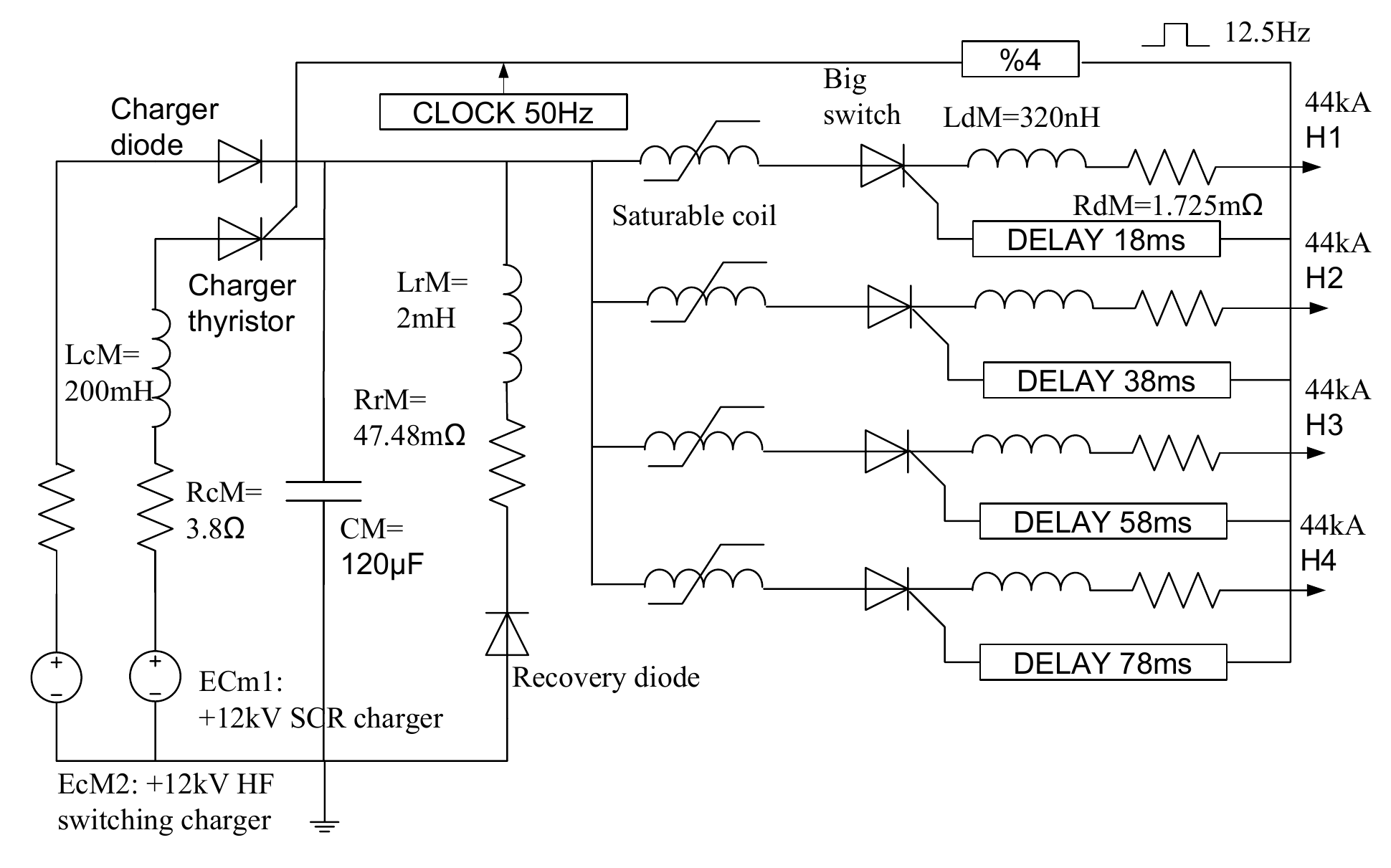}
\caption{\label{PSU_Horns_figure4} Diagram of a power supply module.}
\end{center}
\end{figure}

We have studied a power supply to provide the $\frac{1}{2}$ sinusoid waveform current to the horn.  A capacitor charged at  +12 kV reference voltage will be  discharged through a large switch in a horn via a direct coupled design (Fig.~\ref{PSU_Horns_figure4}). 
A recovery stage allows to invert rapidly the negative voltage of capacitor after the discharge, and to limit the charge capacitor current.

\begin{figure}[htbp]
\begin{center}
\includegraphics[width=11.cm]{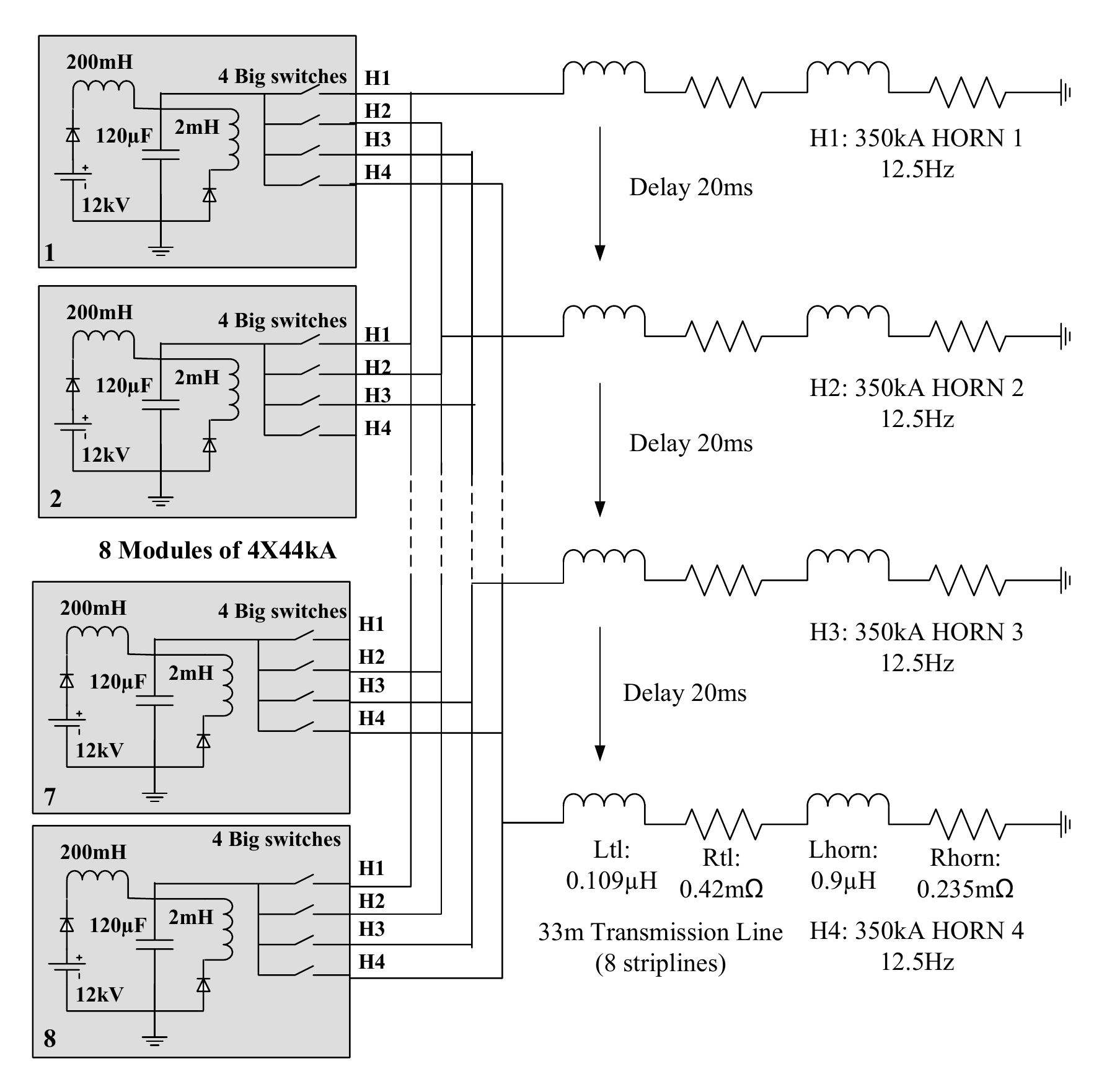}
\caption{\label{PSU_Horns_figure2} Modular architecture of the horns power supply.}
\label{fig:hornps1}
\end{center}
\end{figure}

We have adopted a modular architecture with 8 units (Fig. \ref{fig:hornps1}):  2 modules are interconnected on a same transmission line based on 2 strip-lines (Rtl=1.683 m$\Omega$ and Ltl=435 nH). To limit the energy consumption and therefore the current delivered by the 12kV capacitor charger, investigations have been done to reduce the resistivity and the inductance by studying a transmission line based on large strip-lines of aluminium. It allows to obtain a small resistivity of 51 $\mu \Omega$/m and 13.2 nH/m for 2 plates (0.6 m high X 1cm width) and spaced by 1 cm. 

 The capacitor charge and recovery circuits operate at 50 Hz, the discharge of current in each horn occurs at a 12.5 Hz frequency and is delayed by 20ms between each horn.  

\begin{figure}[htbp]
\begin{center}
\includegraphics[width=8.cm]{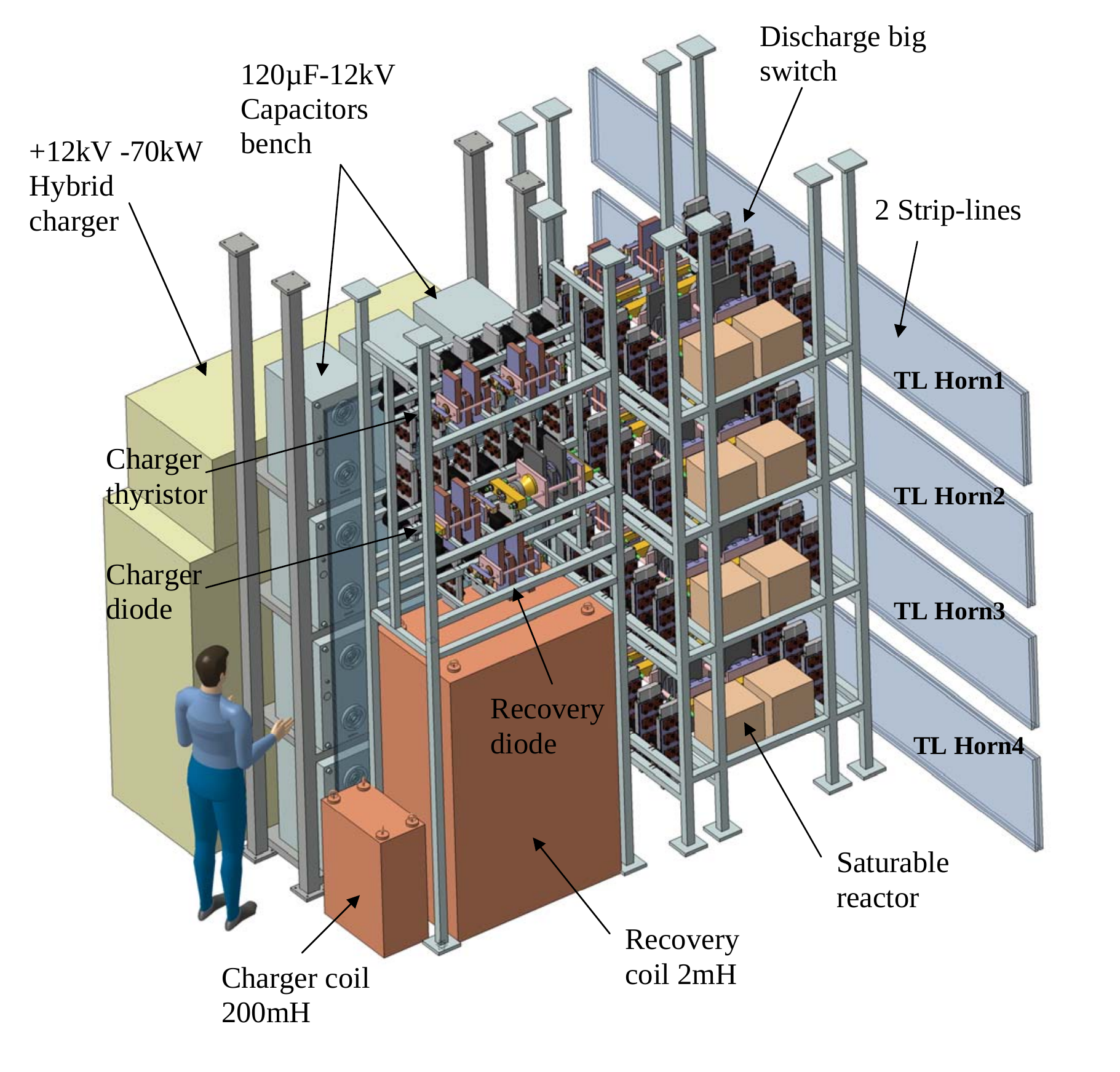}
\caption{\label{PSU_Horns_figure8} Sketch of a power supply unit.}
\end{center}
\end{figure}

The power delivered by the capacitor charger attains 70.8 kW rms per module, that is 566 kW rms in total. It represents only 3\% of quantity of current discharged in horn, so the recovery energy efficiency is very high (97\%).
A sketch of one unit is shown in Fig.~\ref{PSU_Horns_figure8}. A more detailed description of this device can be found in \cite{ref:designreportPS7} and \cite{ref:wp2final}.

\subsection{The target and horn support structure}

\begin{figure}[htbp]
\begin{tabular}{cc}
\includegraphics*[width=7cm]{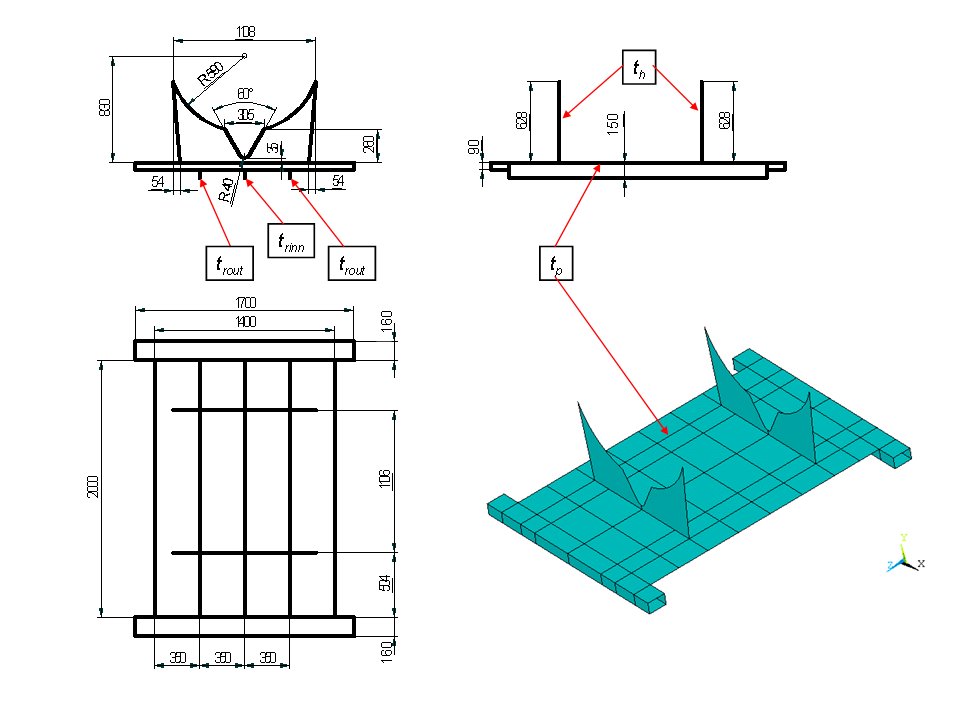} & 
\includegraphics*[width=7cm]{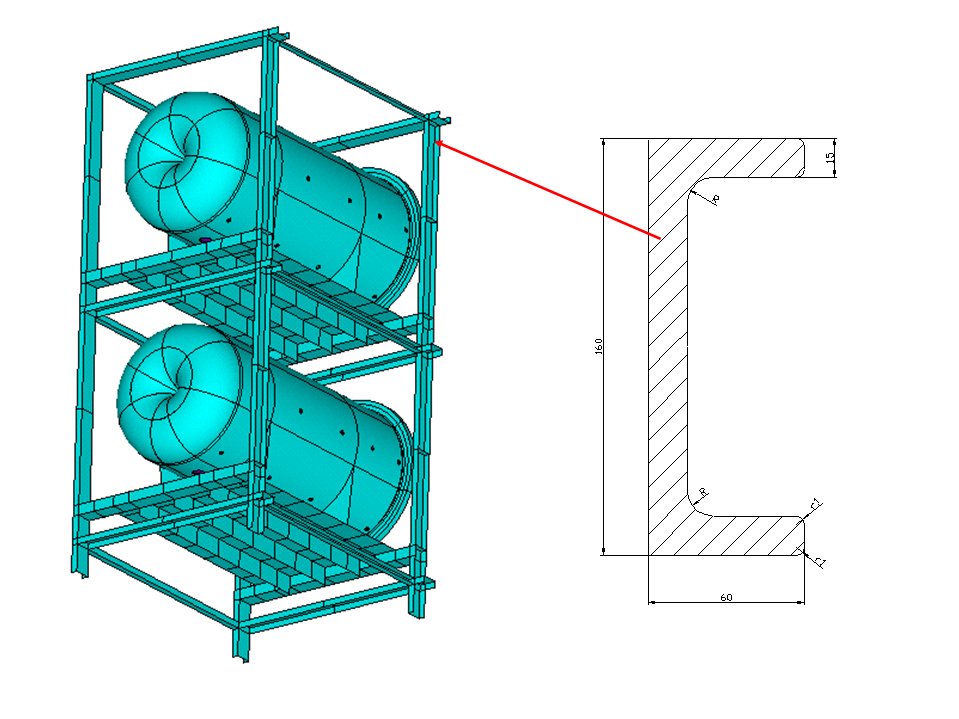} \\
\end{tabular}
\caption{The details of the support for a single horn (left) and 
symmetric half of four horn assembly with detail of channel section used for supporting frame 
 (right). } \label{fig:hornsupport}
\end{figure}

Following the proposal of four horns assembly a supporting structure for the targets and horns has been proposed. This structure consists of a double-sided frame joined with a system of plates directly supporting the horns (Fig. \ref{fig:hornsupport}). The standard channel section was proposed for the frame system (Fig. \ref{fig:hornsupport}).

The thicknesses of the plate elements and reinforcing ribs are proposed on the base of the numerical optimization results, which was performed for the finite element model of the structure. The minimization of the horns deflections was the main optimality criteria used in calculations. In parallel the maximum stress in the horns and the supporting structure were monitored. In the next step the dynamic analysis for the optimized supporting structure with the horns was performed in order to check whether the natural and excitation frequencies are far enough distant, which protects from the resonance. The above described procedure has been performed for two materials used for the supporting system, namely the aluminum alloy (the same as used for the horns) and the construction steel. More detailed results can be found in \cite{ref:wp2final}.

\section{Studies of activation and shielding 	}
\label{sec:activation}

\subsection{Simulation technique}

A detailed calculation of the target and horn activation has been realized with
FLUKA~\cite{manual:fluka,flair} version 2011.2.7 
in order to study the activation of the target and horn and to determine the thickness of shielding required to comply with the radiological regulations. 

The calculation have been done by considering 200 days of irradiation with a 4.5~GeV proton beam of 1MW intensity 
impinging a solid target. The packed-bed target with titanium spheres chosen as the baseline target option is modelled 
as a continuous media with a reduced density of 3 gr/cm$^{3}$~\cite{designreport}. 

%
%
\subsection{Target and horn studies.}

The packed-bed target is placed inside the upstream part of horn's inner conductor and will be represented in the simulation as a 
cylinder 78~cm long with radius 1.5~cm.

\subsubsection{Induced Activation}
The evolution of the induced activation has been estimated as function of cooling time for the target and the horn. The value of the 
specific activity is obtained as a mean value over the total mass of the considered element. 
\begin{figure}[htbp]
\begin{tabular}{cc}
\includegraphics[width=7.5cm]{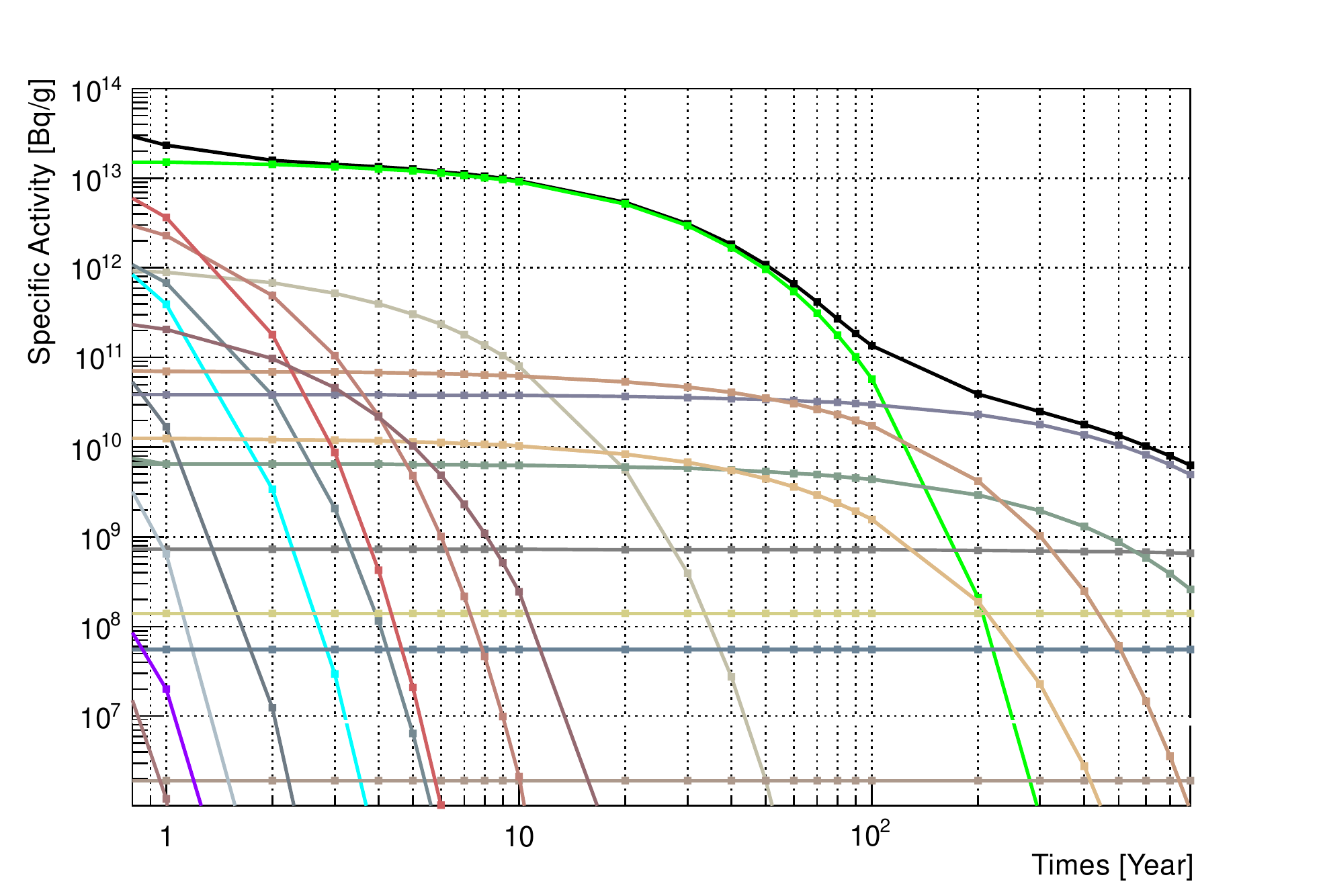} &
\includegraphics[width=7.5cm]{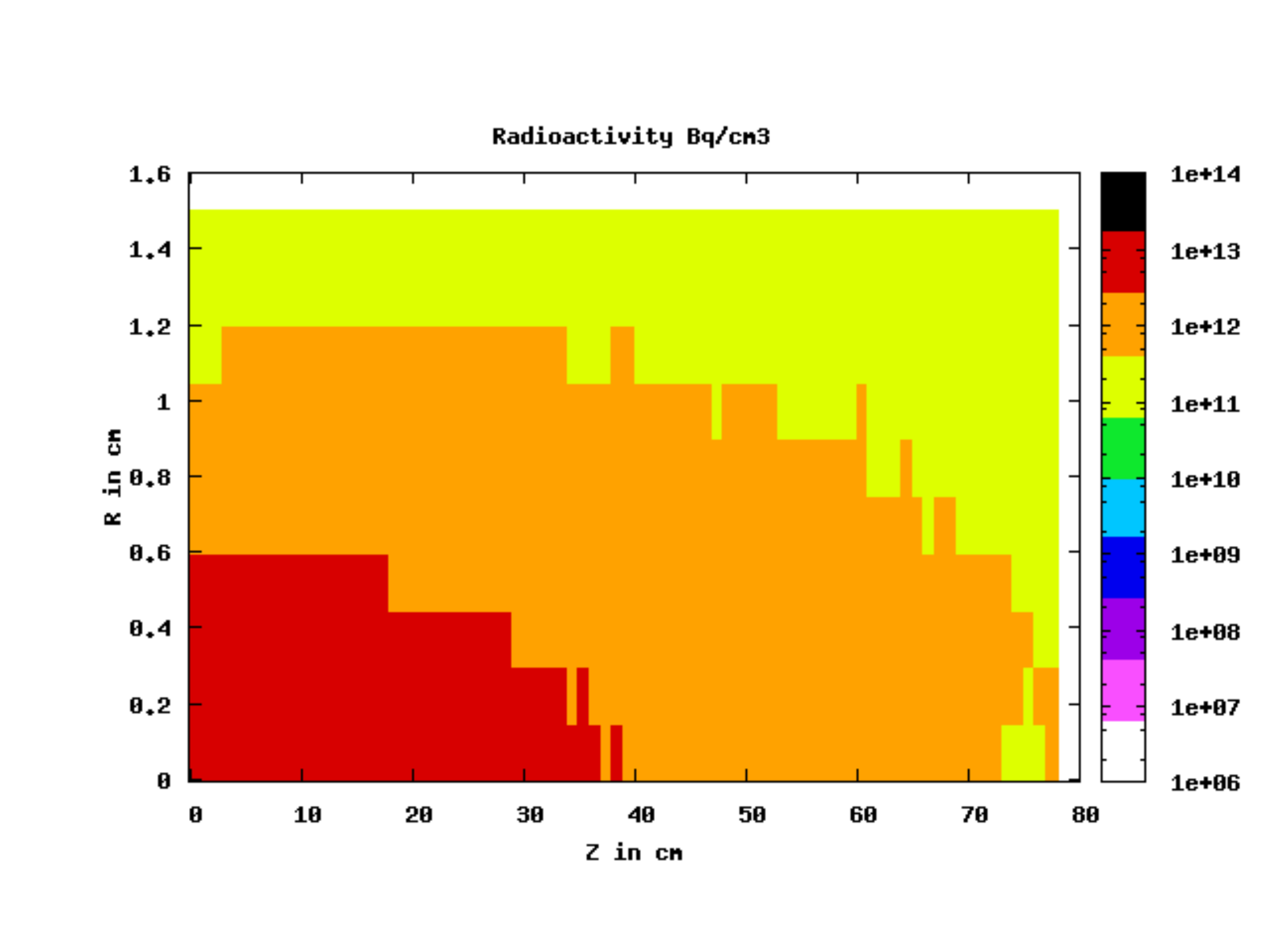} \\
\footnotesize{a) Evolution of the specific activity with cooling times.} &
\footnotesize{b) Spatial distribution of activation in the target.}
\end{tabular}
\caption{ \label{fig:TargetAct} Specific activity of the target.}
\end{figure}
~\\
The activation of the target is non-uniform and present the most active part upstream of the target. The profile of the activation 
follows the energy deposition inside the target with respect to the beam profile (Fig. \ref{fig:TargetAct}). 
\begin{figure}[htbp]
\begin{tabular}{cc}
\includegraphics[width=7.cm]{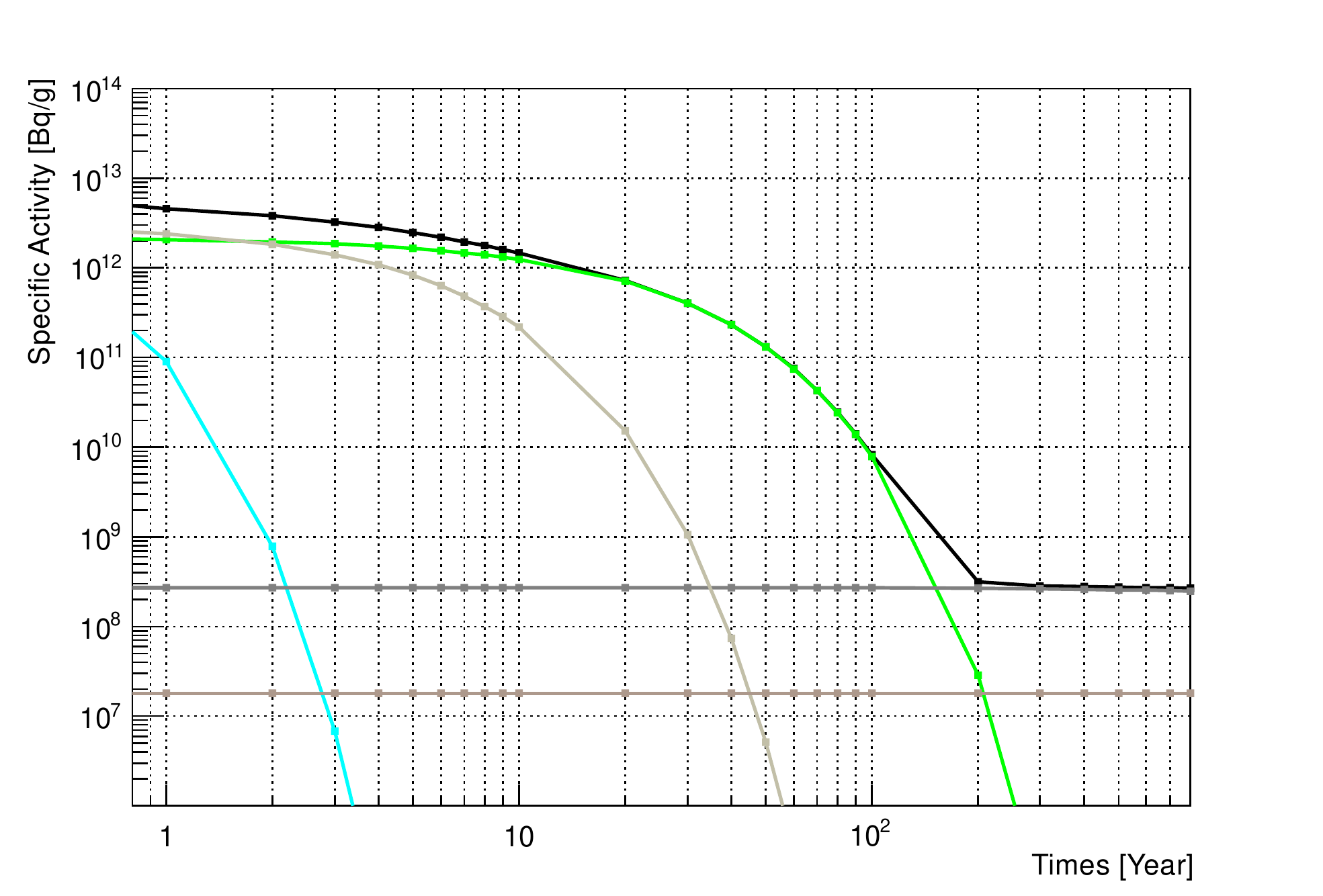} &
\includegraphics[width=7.cm]{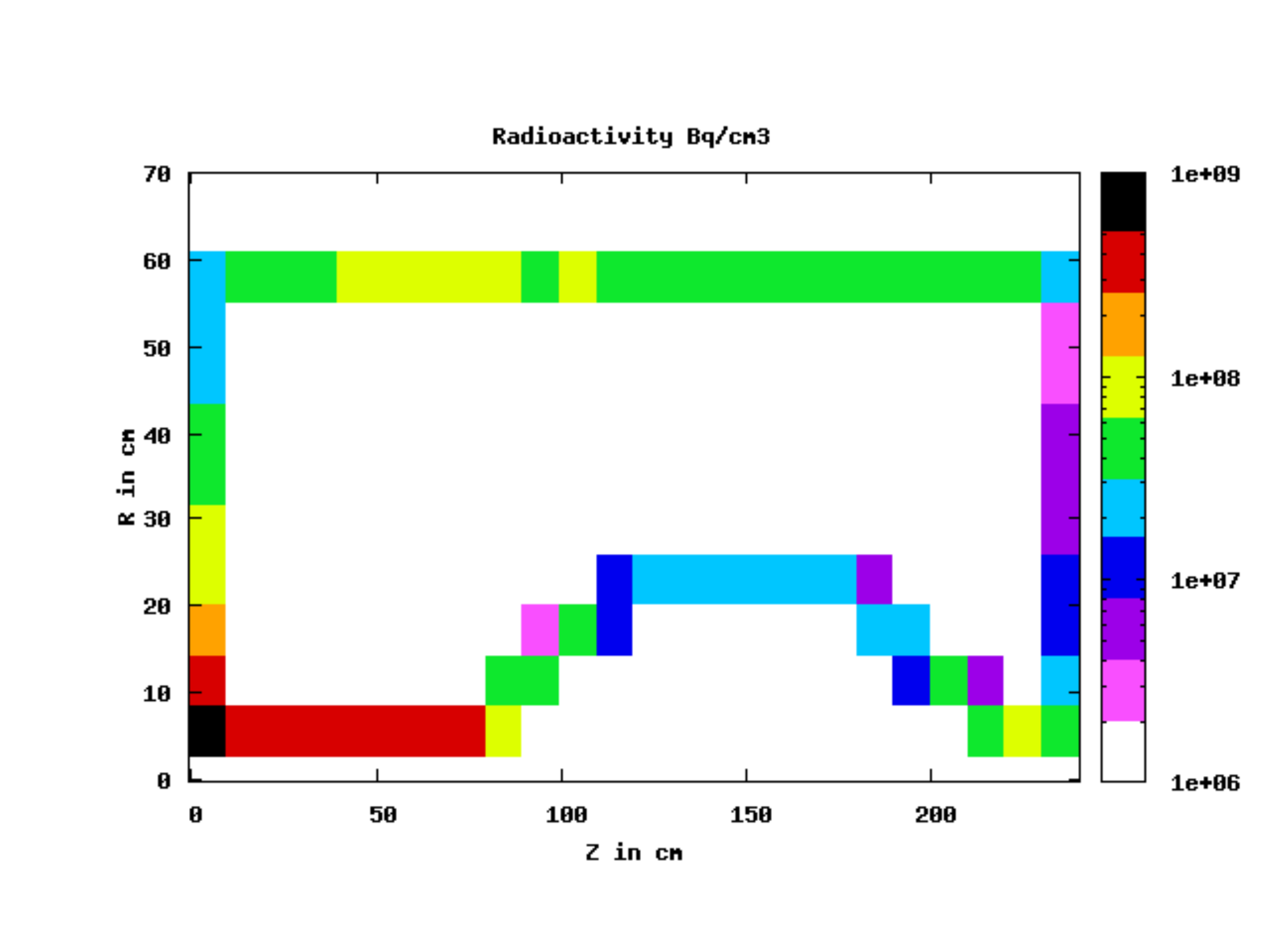}  \\
\footnotesize{a) Specific Activity with cooling times.} &
\footnotesize{b) Spatial distribution of activation.}
\end{tabular}
\caption{ \label{fig:HornAct}  Specific activity of the horn.}
\end{figure}
\\
After one year of cooling time, the remaining radionucleides contributing to the total activity of the horn are $^3$H, 
$^7$Be, $^{10}$B, $^{14}$C, $^{22}$Na and $^{26}$Al  (Fig. \ref{fig:HornAct}) but only gamma emitters have a significant impact on the radiological aspect especially in the case of $^7$Be, $^{22}$Na and the long-lived isotopes $^{26}$Al. 

As in the case of the titanium target, the activation in not uniform inside the horn and presents the most active region in the inner conductor 
as expected (Fig. \ref{fig:HornAct}b). Precautions have to be taken in the building of this part of the inner conductor to prevent cracks 
due to the amount of radiation (water leaks...)

\subsubsection{Dose Equivalent Rate}

A simplified simulation has been realized to evaluate the contribution to the ambient dose 
rate around of the target and the horn thanks to the AMB74 option of FLUKA~\cite{dercoef:fluka}.
 In this study, a two step method has been used to 
evaluate the contribution of each of the elements~\cite{communication:Roesler}.
In this simplified model, all the elements contribute to the dose rate at a non negligeable level. The vessel has an important contribution. 
The concrete has the lowest contribution to the dose rate but the vessel acts as a thin shield in the evaluation.
After one year of operation, the contribution of the horn is still high at the level of 1 Sv/h which prevents human intervention even by removing 
the target which is the most active part by two orders of magnitude compared to the horn. 

%
%
\subsection{Superbeam Facility}
The design of the superbeam facility take advantage of other experiments working with high intensity 
proton beam such CNGS and T2K experiments. 


Secondary charged particles coming out from the pulsed horn will go through the surrounding horns. 
The total energy deposited on them is less than 10 \% of the pulsed horn. 

\subsubsection{Surrounding iron \label{sec:Edep:htggal} }

The simulated geometry and the power densities of the surrounding 
iron and concrete of the four-horn area are presented in Table \ref{tab:station}. 
Results are presented for both neutrino and anti-neutrino beams. A small increase in energy deposition for the anti-neutrino beam is 
due to positive pions de-focusing: more positive secondary particles are produced due to proton-beam charge. Minimal energy deposition 
is seen on the concrete after the iron. The iron vessel and the shield will be cooled with water pipes.

\begin{table}[htbp]
 \centering
  \begin{tabular}{ | c | c | c | c | c | }
    \hline 
     horns-targets area   & target & horn        &  iron              & concrete \\ 
     length = 7.1 m       &        &             &  t = 2.2 m         & t (above horns) = 1.5 m   \\               
                          &        &             & x=-4.9 $->$ 4.9 m  & t (surrounding) = 3.1 m   \\ 
                          &        &             &                    & x = -8 $->$ 8 m           \\ \hline \hline
     v beam (kW)          & 85     &  32         &  437               & 0.01                      \\ \hline
     anti-v beam (kW)     & 85     &  32         &  496               & 0.01                      \\ \hline
  \end{tabular}
  \caption{ Energy deposition  in kW for the horns, iron, and concrete around four-horn system for 4 MW proton beam. In this calculation, an outer conductor and upstream plate thickness of 10 mm has been considered for the horn.\label{tab:station} }
\end{table}

\subsubsection{Decay tunnel \label{sec:Edep:DT} }

\begin{figure}[htbp]
\begin{center}
\setlength\fboxsep{0.5pt}
\setlength\fboxrule{0.5pt}
\fbox{ 
 \includegraphics[width=0.8\textwidth]{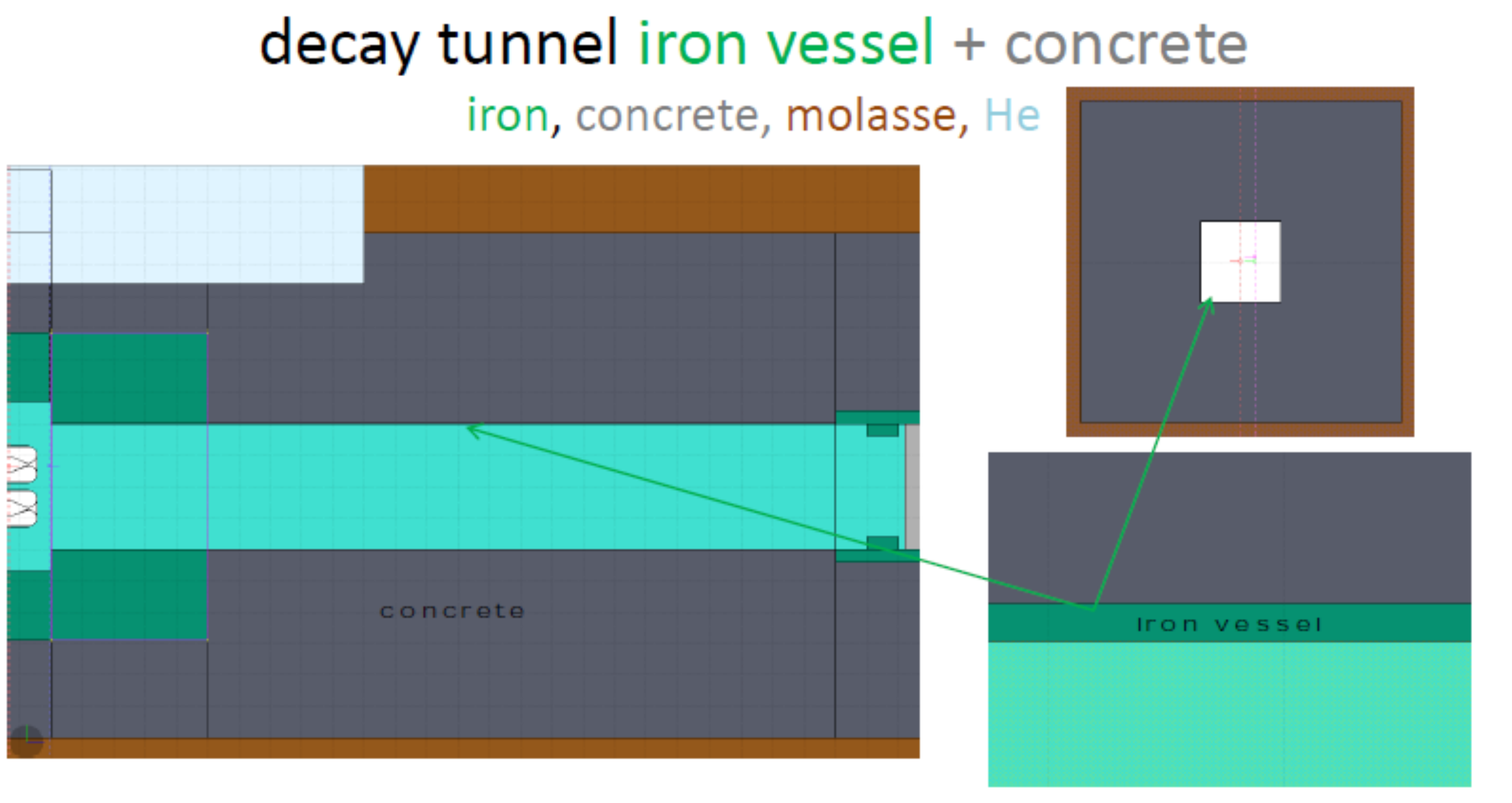}
}
\end{center}
\caption{ Decay Tunnel layout\label{fig:dt}.}
\end{figure}

The decay tunnel area (Fig.~\ref{fig:dt}) consists of the main iron vessel where the particles decay and neutrinos are produced, and the concrete surroundings in order to protect the molasse from activation. At the beginning of the decay tunnel an upstream iron-shield is also foreseen to protect the areas above it like the strip-lines. The horn power supply will be built above the start of the decay tunnel.  

The energy deposition and the dimensions and thicknesses for the decay-tunnel iron vessel, concrete and upstream iron collimator are shown in Table \ref{tab:dt1} and Table \ref{tab:dt2}. That geometry is optimized to keep the activation at minimum level in molasse. 
The decay tunnel vessel will be cooled by water pipes.

\begin{table}[htbp]
 \centering
  \begin{tabular}{ | c | c | c | }
    \hline 
     area                  & DT iron vessel        &  DT surrounding concrete  \\ 
     length = 25 m         & H, W = 4 m           &  t = 6 m                 \\               
                           & t = 1.6 cm    &                          \\ \hline \hline
     v beam (kW)           &  390                  &  485                    \\ \hline
     anti-v beam (kW)      &  392                  &  588                    \\ \hline
  \end{tabular}
  \caption{ Energy deposition in kW for the decay tunnel iron vessel and surrounding concrete\label{tab:dt1}. }
\end{table}

\begin{table}[htbp]
 \centering
  \begin{tabular}{ | c | c | c | }
    \hline 
     area                  & DT iron shield        &  DT iron shield-above decay tunnel  \\ 
     length = 25 m         & t = 2.9 m             &  t = 2.9                            \\               
                           & x = -4.9 $->$ 4.9 m   &                                     \\ \hline \hline
     v beam (kW)           &  610                  &  159                                \\ \hline
     anti-v beam (kW)      &  775                  &  201                                \\ \hline
  \end{tabular}
  \caption{ Energy deposition in kW for the decay tunnel upstream iron shield \label{tab:dt2}.}
\end{table}

Preliminary calculation within the WP2 group show that water-cooling is feasible for the decay tunnel vessel. 

\subsubsection{Beam dump \label{sec:Edep:BD} } 

\begin{figure}[htbp]
\begin{center}
\setlength\fboxsep{0.5pt}
\setlength\fboxrule{0.5pt}
\fbox{ 
 \includegraphics[width=0.6\textwidth]{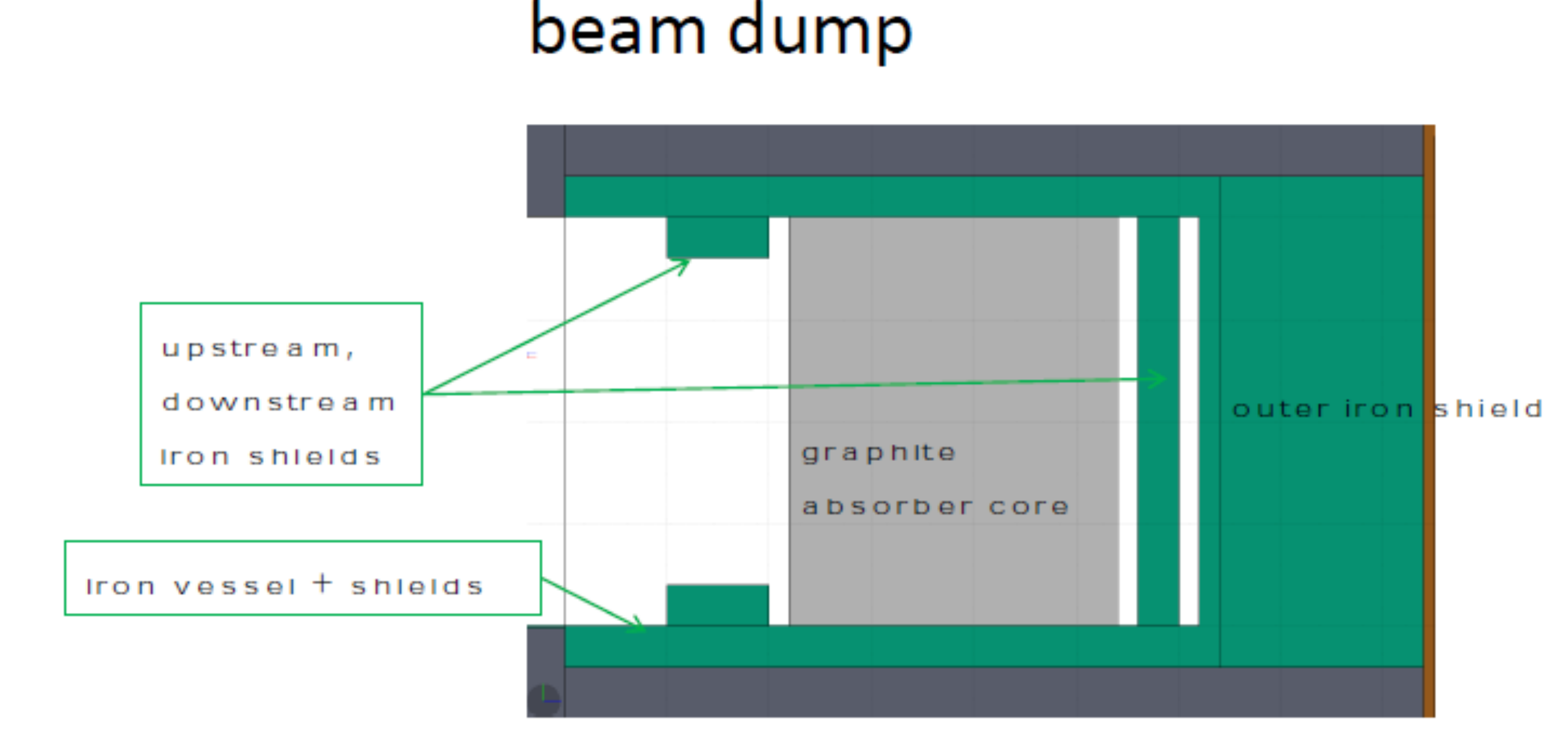}
} 
\end{center}
\caption{ Beam dump layout used in simulation. Graphite beam dump in grey and several iron shields in green\label{fig:bd1}.}
\end{figure}


The beam dump area for the SPL Super Beam follows the design of the T2K \cite{ref:t2ktarget}. It consists of the main graphite block and several shields (\ref{fig:bd1})
with the purpose to dump the remaining hadron particles and finally confine the hadronic energy within the experimental layout.

\begin{table}[htbp]
 \centering
  \begin{tabular}{ | c | c | c | c | c | c | }
    \hline 
         area         & graphite     &  up-shield         & down-shield              & outer shield & surrounding \\ 
     length = 6.4 m   & L = 3.2 m    &                    &                          &              & concrete    \\               
                      & H, W = 4 m   &                    &                          &              & t = 6 m     \\ \hline \hline
     v beam (kW)      &  778         &  146               & 19                        & 1.6          & 4           \\ \hline
     anti-v beam (kW) &  485         &  128               & 12                        & 1            & 3.6         \\ \hline
  \end{tabular}
  \caption{ Total energy deposition in kW for the graphite beam dump and various shields.\label{tab:bd}. }
\end{table}

The energy deposition values are shown in Table \ref{tab:bd}.
The beam dump absorbs all the remaining hadrons so the activation of 
molasse or any other installation after the beam is prevented.  As a result of that, high power dissipation is developed on the dump. On-going studies 
show the graphite beam dump operation will be feasible by using helium conduction along gaps in graphite. Additional studies show that the induced radioactivity 
in molasse is kept well under the CERN's limits \cite{tritium}. 

%
%
\subsubsection{Summary \label{sec:summary} }

\begin{figure}[htbp]
\begin{center}
\includegraphics[angle=0, width=10cm]{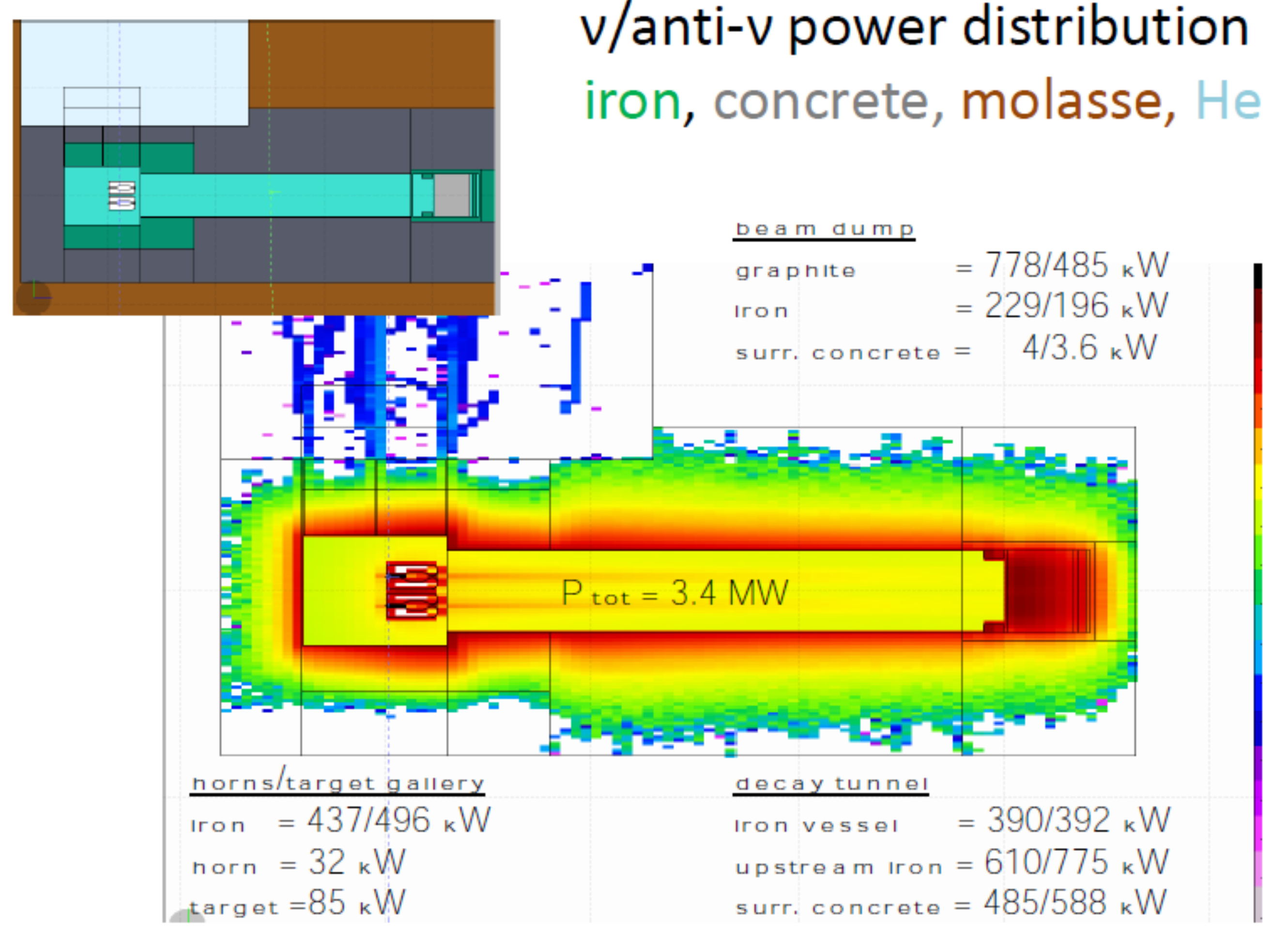}
\end{center}
\caption{ Summary of the power densities for the neutrino and anti-neutrino Super Beams\label{fig:all}.}
\end{figure}

With this simulation, we have studied the power dissipation on different elements of the Super Beam. The summary for the neutrino and anti-neutrino beam is presented in Fig. \ref{fig:all}. These data are used as input to the finite-elements calculations for the heat dissipation and 
the design of the cooling methods for the titanium target, the aluminum horn and the graphite beam dump \cite{designreport}.

%
%


%
%
\subsubsection{Shielding investigation}
A first approach for the estimation of the shielding is based on a geometry consisting of a simple iron layer surrounded by concrete. 
The prompt dose rate can be estimated by using an empirical formulae giving the attenuation:
$$
H=\frac{H(\theta)}{R^2}.e^{-\frac{t}{\lambda}}
$$
with t the total thickness of the material and $\lambda$ the equivalent length in iron and concrete. 

If the design of the structure of the shielding element is kept similar to the T2K, by considering the 2.2m of iron, the concrete
thichness should reach 3.7 m  to decrease the prompt dose rate at a level of $10~\mu Sv$.

\section{Optimization, fluxes and sensitivity}
\label{sec:optimization}
\subsection{Physics performances}

This section summarizes the main results appearing in \cite{ref:andreaPub}.
The neutrino energy spectra are calculated using a probabilistic approach in
order to obtain reliable results in a reasonable amount of time using samples 
of $\sim$ $10^6$ simulated protons.
The probability that the neutrino will reach the far detector is calculated at each
particle decay yielding neutrinos with analytic formulas
\cite{Campagne:2004wt,Campagne:2006yx,Blondel:2000ph,Cazes:1900zz}. The probability is then 
used as a weight factor in the calculation of the neutrino energy spectrum.
Neutrinos from hadron interactions in the walls of the decay tunnel or in 
the beam dump are neglected in this approach.
 
The distribution of the secondaries at target exit obtained with the FLUKA \cite{Battistoni:2007zzb} generator
is used as an external input to a GEANT4 \cite{GEANT4} simulation derived from a  GEANT3 code developed in \cite{Campagne:2004wt}.
The target, the horn with its magnetic field and the decay tunnel are fully simulated 
within GEANT4. Alternatively GEANT4 can be used to simulate also the interactions of primary protons in the target:
this option was used as cross check.
In order to cross-check and validate the new GEANT4--based software, a
comparison has been done with the fluxes obtained with GEANT3. 
The fluxes obtained in the two frameworks are in good agreement both in terms of normalization and shape \cite{NUFACT09}.
Further cross-checks included the correct implementation of the decay branching ratios,
a comparison with an independent code and a check based on direct scoring of the emitted neutrinos.

The sensitivities for the measurement of the oscillation parameters
$\theta_{13}$ and $\delta_{CP}$ are obtained with the help of GLoBES 3.0.14
\cite{Huber:2004ka}.

\subsection{Target and horn optimization}

The approach which was followed in the optimization of the forward--closed horn and 
the decay tunnel uses the final $\sin^2 2\theta_{13}$ sensitivity. This is a way to maximise the flux at the 
first oscillation maximum.  In this way the final physics performance is used as a
guiding principle in the ranking of the configurations under scrutiny. In the evaluation
of this quantity a complex set of relevant factors are given as an input: the normalization and shape 
of each neutrino flavor, the running time in the positive and negative focusing mode, 
the energy dependence of the cross sections, the backgrounds in the far detector and its response
in terms of efficiency and resolution.

We define the $\delta_{CP}$-dependent 99\% C.L. sensitivity limit as $\lambda_{99}(\delta_{CP})$. Averaging on $\delta_{CP}$ and multiplying by $10^{3}$
we introduce:
\begin{equation}
  \label{eq:lambdadef}
  \lambda 
  =
   \frac{10^3}{2\pi} \int_{0}^{2\pi} 
     \lambda_{99}(\delta_{CP})\, d\delta_{CP}
\end{equation}
This quantity has been used as a practical way of defining with a single number the 
quality of the focusing system. 

The key parameters defining the horn and tunnel geometry are randomly sampled within specified 
ranges and the correlations with the figure of merit $\lambda$ studied. 

The parameters of the forward--closed horn and of the decay tunnel were sampled with uniform 
probability distributions imposing the configuration to be  geometrically consistent 
(``iteration-1'').  After studying the correlation of these parameters with the figure of merit, a second iteration was performed with a restriction of the phase spaces around the most promising values. The geometrical parameters obtained with this optimization have been reported previously in this article. 

\subsection{Beam fluxes}

\begin{figure}[htbp]
\begin{center}
\includegraphics[width=10cm]{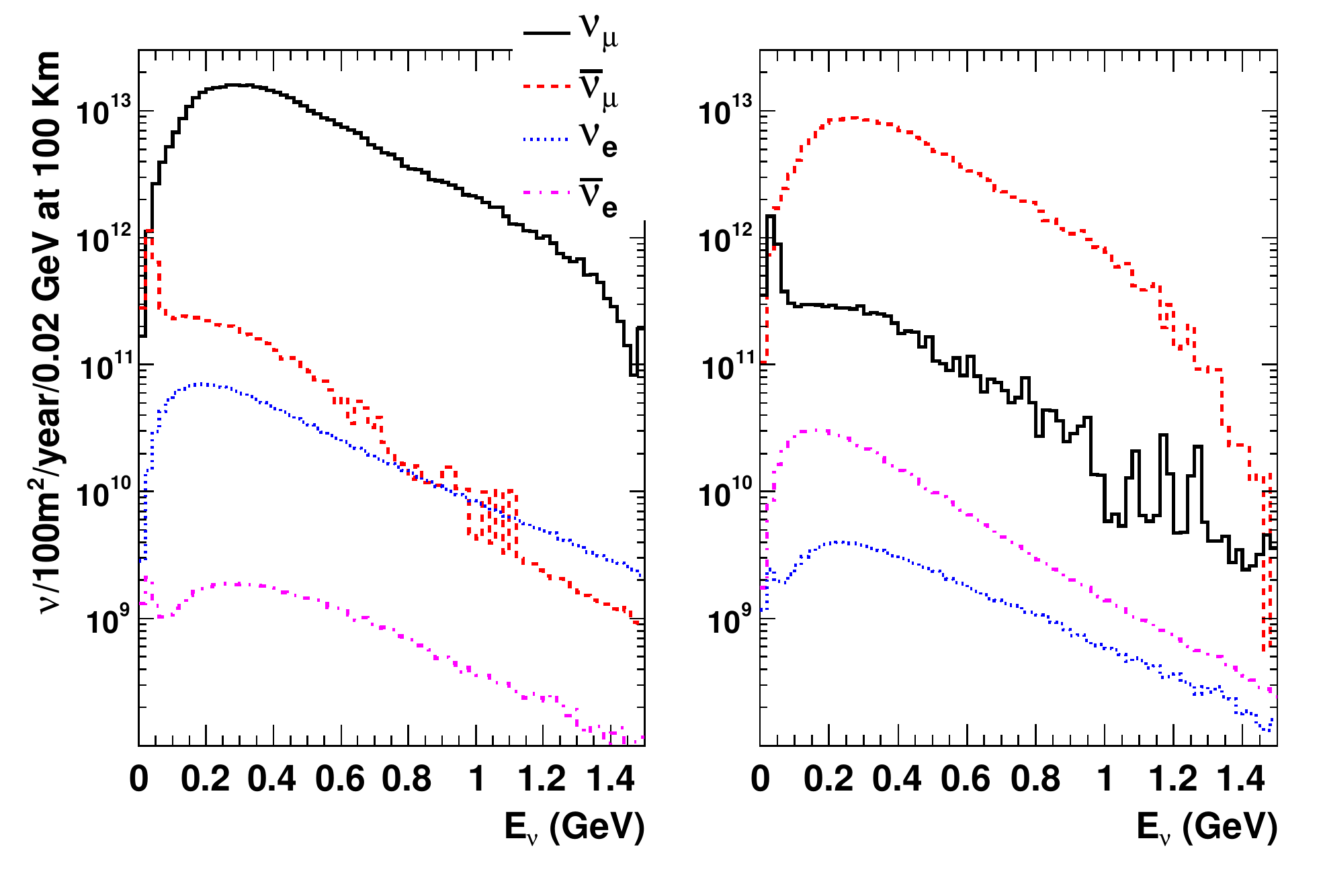}
\caption{Non-oscillated neutrino fluxes obtained with the optimized horn and decay tunnel
 in positive (left) and negative (right) focusing mode. }
\label{fig:fluxesFC}
\end{center}
\end{figure}

The non-oscillated $\nu_\mu$, $\nu_e$ and charged conjugate (c.c.) neutrino fluxes 
are shown in Fig.~\ref{fig:fluxesFC} for positive (left) and negative focusing (right) runs. 
They correspond to 5.6~$\cdot$~10$^{22}$ protons on target (p.o.t.)/year
(4~MW $\cdot$ 10$^7$ s at 4.5~GeV) and are calculated at a reference distance of 100~km over a surface of 100~m$^2$. 
The fractions of $\nu_\mu$, $\bar{\nu}_\mu$, $\nu_e$ and $\bar{\nu}_e$ with respect to the total
are (98.0\%, 1.6\%, 0.42\%, 0.015\%) and (4.4\%, 95.3\%, 0.05\%, 0.28\%) for the positive and negative focusing 
modes respectively.

\begin{table}[htbp]
\centering
\begin{tabular}{|c|c|c|c|c|}
\hline
focusing &$\nu_\mu$ & $\bar{\nu}_\mu$ & $\nu_e$ & $\bar{\nu}_e$\\ 
\hline
+ & $3.9\cdot 10^{14}$ & $6.3\cdot 10^{12}$ & $1.7\cdot 10 ^{12}$ & $6.0\cdot 10^{10}$ \\
&\scriptsize{98.0\%}& \scriptsize{1.6\%}& \scriptsize{0.42\%}& \scriptsize{0.015\%}\\
\hline
- & $1.0\cdot 10^{13}$ & $2.2\cdot 10^{14}$ & $1.2\cdot 10 ^{11}$ & $6.4\cdot 10^{11}$ \\
&\scriptsize{4.4\%} & \scriptsize{95.3\%} &\scriptsize{0.05\%} &\scriptsize{0.28\%}\\
\hline
\end{tabular}
\caption{ Integral neutrino flux per year for each flavor at 
a distance of 100~km over a surface of 100~m$^2$. The fluxes were obtained with a sample of $10^7$ simulated proton-target interactions.}
\label{tab:fluxesFC}
\end{table}

In positive (negative) focusing mode the $\nu_e$ ($\bar{\nu}_e$) 
fluxes are dominated by muon decays: 82\% (90\%). 
The c.c. fluxes receive instead a large contribution from kaon three body decays 
(81\% and 75\% in positive and negative focusing respectively)
with muon decays from the decay chain of ``wronge charge'' pions at low energy
contributing for the rest. The fluxes are publicly available \cite{fluxesWEB}.

The fluxes obtained with the optimized horn have been compared to those obtained with 
the original double conical horn with currents of 300 and 600~kA associated with a mercury target and published in \cite{Campagne:2006yx}. 
The $\nu_\mu$ and $\nu_e$ energy spectra are shifted to higher
energies with an increase in statistics particularly around 500
MeV. The $\nu_\mu$ flux is enhanced also in the proximity of the oscillation maximum at 260 MeV 
where  the $\nu_e$ flux is reduced by a similar fraction. The wrong-CP component ($\bar{\nu}_e$, $\bar{\nu_\mu}$) 
on the other hand is reduced by more than a factor two. 

\begin{figure}[htbp]
\begin{center}
\includegraphics[width=10cm]{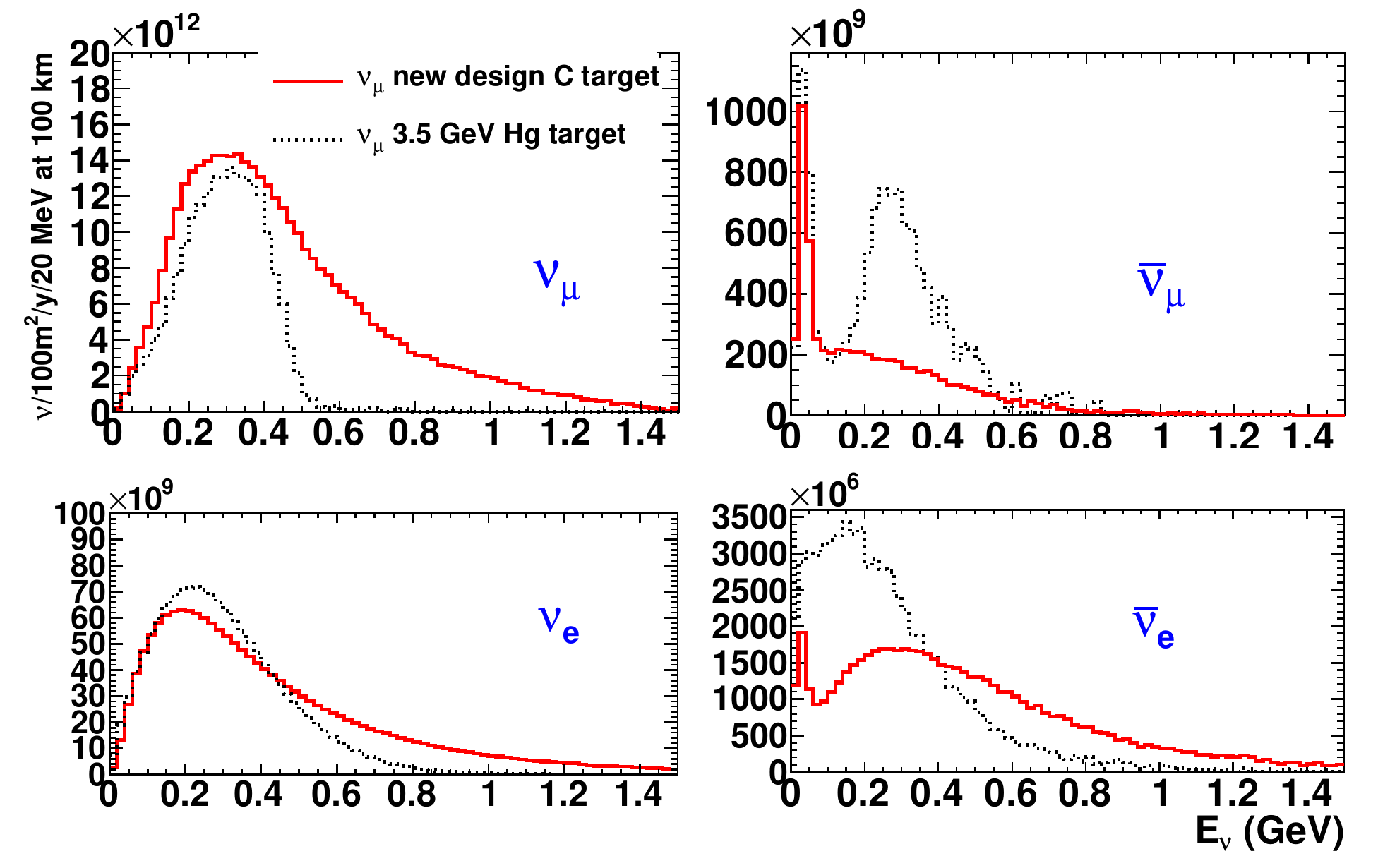}
\caption{Comparison of the neutrino fluxes obtained with the new design (continuous line) and the previous one \cite{Campagne:2006yx} (dotted line) }
\label{fig:fluxesNEWOLD}
\end{center}
\end{figure}

\subsection{Physics performances}

\begin{figure}[htbp]
\begin{center}
\includegraphics[width=10cm]{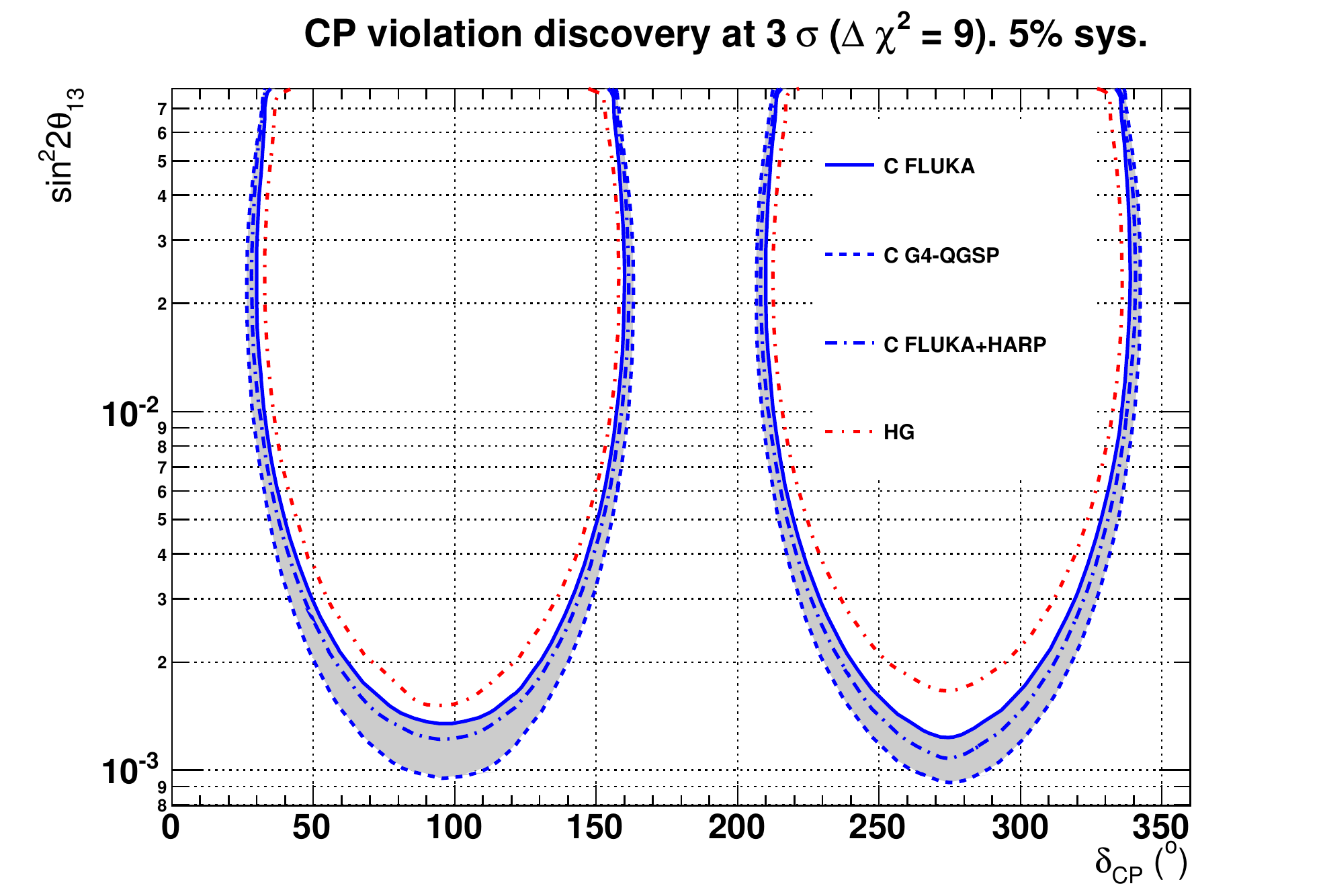}
\caption{CP violation discovery potential at 3~$\sigma$ level. See text.}
\label{fig:CPVb}
\end{center}
\end{figure}

The CPV discovery potential at the 3~$\sigma$ level is shown in Fig.~\ref{fig:CPVb}: discovery is possible in the region 
above the curves. This means that in that region of the true ($\sin^22\theta_{13}$, $\delta_{CP}$) plane a fit done
under the CP conserving hypotheses ($\delta_{CP}=0, \pi$) gives for both choices
a $\Delta\chi^2>9$. 
The limit obtained with the previous setup associated with the mercury target is shown by the dash-dotted curve while the new limits are represented as a hatched band. 
The upper edge of the band (continuous line) refers to the FLUKA model of hadro-production, the lower edge (dotted) to the GEANT4-QGSP model, the one lying (mostly) in 
the middle (long dash-dotted) is obtained after reweighting FLUKA with the HARP data. 
The new limits generally improve those obtained with the previous design both for $\theta_{13}$ and CPV 
discovery.

 \section{Conclusions}
\label{sec:conc}
This study is the first that presents a clear and complete conceptual design for a very challenging facility, capable of delivering a low energy neutrino beam with a 4 MW 4.5 GeV/c proton driver. We have presented a novel design for the target, using both a split proton beam to divide the power on each device by a factor four and a pebble bed target. The latter allows the coolant to dissipate in a very efficient way the heat, flowing through the innermost part of the target. The structure of the Ti spheres is such that they will stand the static and dynamic stresses. Preliminary calculations show that this target will be able to stand not only 1 MW per device, as originally required, but probably a higher power. This feature makes it a very attractive solution also for other facilities.

The focusing device, a magnetic horn, based on a conventional design, has been optimized for our needs on the basis of new approach that allow to study a large parameter space, defined by its geometry, material thickness, current and the decay tunnel characteristics. This optimization has allowed to maintain the excellent physics performances while offering a realistic design. Preliminary studies conclude that the lifetime of each device will be sufficient for a routine operation with high reliability. A difficult but key component is the power supply, subject to an unusual high repetition rate of 50 Hz for a peak current of 300 kA. 

We have studied most of the system features, starting from the proton beam exiting from the accumulator up to the beam dump. This has required a diverse array of complementary competences and studies which are only briefly summarized here. Our main conclusion is that this project is feasible by adopting the novel approach that we have introduced and developed here. We have fully studied the shielding and activation issues, to comply with existing radiological regulations, and found that the shielding type and thicknesses, while sizeable, are not excessive neither in terms of engineering nor of cost. In general, while some of the problems that we had in front of us at the start of the project were particularly challenging, we have found no show-stopper and are confident that this project is feasible. 

Of course, this study, developed within the context of EUROnu, was limited to the engineering and simulations levels. Some of the devices considered here are novel and would require an extensive phase of R\&D to assess their performances and validate with a prototype their use in this context.

\section*{Acknowledgments}

We acknowledge the financial support of the European Community under the
European Commission Framework Programme 7 Design Study: EUROnu, Project
Number 212372. The EC is not liable for any use that may be made of the
information contained herein.





\end{document}